\documentclass[sigconf, arxiv]{acmart}
\usepackage{multirow}
\usepackage{graphicx}
\usepackage{tabularx} % Add this to use tabularx
\usepackage{booktabs} % For professional looking tables
\usepackage{array}
\usepackage{geometry}
\usepackage{adjustbox}
\usepackage{enumitem}
\usepackage{pdfpages}
\setlist[itemize]{leftmargin=*}

\newcommand{\sys}{{SPHERE}{}}

\AtBeginDocument{%
  }

\setcopyright{acmlicensed}
\copyrightyear{2018}
\acmYear{2018}
\acmDOI{XXXXXXX.XXXXXXX}

\acmConference[Conference acronym 'XX]{Make sure to enter the correct
  conference title from your rights confirmation emai}{June 03--05,
  2018}{Woodstock, NY}
\acmISBN{978-1-4503-XXXX-X/18/06}
\begin{document}

\title{\sys{}: Scaling Personalized Feedback in Programming Classrooms with Structured Review of LLM Outputs}

\author{Xiaohang Tang}
\affiliation{%
  \institution{Virginia Tech}
  \city{Blacksburg}
  \state{Virginia}
  \country{USA}
}
\email{xiaohangtang@vt.edu}

\author{Sam Wong}
\affiliation{%
  \institution{University of Washington}
  \city{Seattle}
  \state{Washington}
  \country{USA}
}
\email{samw627@uw.edu}

\author{Marcus Huynh}
\affiliation{%
  \institution{Virginia Tech}
  \city{Blacksburg}
  \state{Virginia}
  \country{USA}
}
\email{mjhuynh@vt.edu}

\author{Zicheng He}
\affiliation{%
  \institution{ University of Virginia}
  \city{Charlottesville}
  \state{Virginia}
  \country{USA}
}
\email{bgc4bx@virginia.edu}

\author{Yalong Yang}
\affiliation{%
  \institution{Georgia Institute of Technology}
  \city{Atlanta}
  \state{Georgia}
  \country{USA}
}
\email{yalong.yang@gatech.edu}

\author{Yan Chen}
\affiliation{%
  \institution{Virginia Tech}
  \city{Blacksburg}
  \state{Virginia}
  \country{USA}
}
\email{ych@vt.edu}

\renewcommand{\shortauthors}{Xiaohang Tang, Sam Wong, Marcus Huynh, Zicheng He, Yalong Yang, and Yan Chen}

\begin{abstract}
This paper introduces \sys{}, a system that enables instructors to effectively create and review personalized feedback for in-class coding activities. 
Comprehensive personalized feedback is crucial for programming learning. However, providing such feedback in large programming classrooms poses significant challenges for instructors. 
While Large Language Models (LLMs) offer potential assistance, how to efficiently ensure the quality of LLM-generated feedback remains an open question.
\sys{} guides instructors' attention to critical students' issues, empowers them with guided control over LLM-generated feedback, and provides visual scaffolding to facilitate verification of feedback quality.
Our between-subject study with 20 participants demonstrates \sys{}'s effectiveness in creating more high-quality feedback while not increasing the time spent on the overall review process compared to a baseline system. This work contributes a synergistic approach to scaling personalized feedback in programming education, addressing the challenges of real-time response, issue prioritization, and large-scale personalization.

\end{abstract}

% \keywords{Do, Not, Us, This, Code, Put, the, Correct, Terms, for,


\maketitle
\section{\MakeUppercase{Introduction}}
When learning to program, personalized feedback is among the most effective ways to help students bridge the gap between their current understanding and performance and their desired learning goals \cite{hattie2007power,butler1995feedback,hattie2012visible}. 
To create effective personalized feedback, instructors need to assess students' coding challenges in the context of class learning objectives, consider each student's competencies and help resources, and tailor guidance to both the specific problem and the student's cognitive state.
However, in programming classrooms, it is challenging to provide effective personalized feedback due to the sheer volume of time-sensitive issues arising from ongoing class exercises, such as coding and small group discussions, far exceeding instructors' attention capacity.

Past research and our own formative work with programming instructors have shown a large gap between the effort they put into providing feedback to students during class exercises using existing methods and the increasing number of questions that students accumulate during the exercises. Whether in-person (e.g., raising hands) or technology-based (e.g., learning dashboards like VizProg \cite{zhang2023vizprog} and VizGroup \cite{tang2024vizgroup}), these methods fall short of addressing this challenge. In fact, we argue that in-person approaches often result in less-personalized feedback due to time constraints and social context during the exercises, while learning dashboards can lead to information overload in large classes. This overload, combined with human and UI biases, often leads instructors to focus on issues that are more familiar or quickly addressable, rather than those that may be most critical. This tendency is driven by the social pressure to progress through the exercise, potentially overlooking more complex or unfamiliar issues that require deeper consideration.
These limitations highlight three critical support needs for instructors: prioritizing critical issues, providing real-time feedback on these issues, and ensuring personalization of feedback at scale. 

In this paper, we reframe this challenge as an attention-direction problem, where instructors aim to leverage system support while maintaining a sense of control over their cognitive efforts.
Past research has attempted to address this challenge, but has focused on only one or two of these aspects.
For instance, Codeoption~\cite{guo2015codeopticon} allows instructors to monitor students' ongoing coding activities, but its design of displaying every student's code changes limits scalability to only a dozen or two students at most. VizProg \cite{zhang2023vizprog}, and VizGroup~\cite{tang2024vizgroup} enable instructors to analyze a large scale of students' code and group discussions in real-time, but they either lack guidance on prioritizing specific sub-issues or left the feedback creation stage untouched.
LA Cockpit introduces a template-based system for feedback creation, but the rule-based approach is difficult to scale to dynamic, ongoing students' activities~\cite{karademir2024LACockpit}.

Recent advances in Large Language Models (LLMs) offer the potential for real-time, context-relevant feedback generation at scale. However, the use of LLMs presents its own challenges. Unlike rule-based or expert-led approaches, LLMs often generate multiple different responses to a single prompt, with variations in both accuracy and level of personalization. This can lead to hallucinations \cite{huang2023surveyhallucinationlargelanguage,Bender2021OnTD} and misalignment with class-specific learning objectives \cite{Sridhar2023HarnessingLI}. Successfully identifying and repairing inaccuracies and misalignments in LLM outputs require domain expertise, necessitating a review by experts, such as instructors, before their adoption for educational use.
Therefore, we ask: \textit{How can we design a system to support instructors in effectively reviewing large-scale LLM-generated feedback on in-class coding activities?} 

In this paper, we present \sys{}\footnote{\sys{} is an acronym for \textbf{S}caling \textbf{P}ersonalized \textbf{HE}lp in \textbf{RE}al-time}, an interactive system designed to help programming instructors create personalized feedback at scale for class exercises, including writing code and discussing code in groups. \sys{} employs a novel approach that combines intelligent issue detection with structured feedback generation and review. Specifically, \sys{} first employs a novel LLM architecture that continuously identifies key patterns in student coding and group activities, and directs instructors' attention based on issue severity. Then, \sys{} uses our ``strategy-detail-verify'' approach to ensure the quality of feedback created: 1) it allows instructors to \textit{strategically} guide the LLM to generate high-quality feedback corresponding to these issues using key feedback components, and 2) it also provides information visualization bindings to facilitate instructors' rapid \textit{verification} of the personalization and accuracy of feedback (\textit{details}) in relation to corresponding issues.

The benefits of \sys{} lay in its ability to support instructors in efficiently creating high-quality, personalized feedback on dynamic students' learning activities at scale. Our key insight is that by templating essential elements of high-quality feedback and integrating visual mapping for verification, instructors can more quickly and effectively create, contextualize, and verify the accuracy and personalization of LLM-generated feedback.

% The core contribution of \sys{} is a novel integration of LLM-powered issue detection and feedback generation with a structured instructor review process. 

To evaluate the effectiveness of \sys{}, we conducted an in-lab, between-subject study with 20 instructors. We compared \sys{} to a baseline system that used a standard linear structure view to present LLM-generated feedback.
% This comparison focused on the efficiency and quality of the feedback creation and review process, rather than the underlying AI model capabilities.
Our findings indicate that \sys{} generated and sent significantly more high-quality feedback ($p < 0.01, 0.01$, respectively) and can help instructors transform more low-quality feedback into high-quality ones more often ($p = 0.001$). Qualitatively, participants reported feeling more thoughtful about the feedback they created for students. Interestingly, we did not observe a significant increase in the time participants spent using \sys{}, indicating that our approach to create feedback has a low net cost from an interaction perspective. 
These results underscore \sys{}'s capacity to both streamline the feedback process and deepen instructors' engagement with feedback creation, which in turn can enable more impactful and aligned student guidance.
Our work makes the following contributions:

\begin{itemize}
    \item Insights into key challenges and opportunities in scaling personalized feedback for in-class programming exercises, revealed by our formative study.
    
    \item A novel method that integrates LLM-powered issue detection and feedback generation with a structured instructor review process.

    \item \sys{}, an interactive system that guides instructors' attention to critical issues in real-time, enables guided control over LLM-generated feedback, and facilitates efficient quality assurance of personalized responses.

    \item Empirical evidence demonstrating the effectiveness of \sys{} in improving feedback quality and instructor efficiency in large-scale programming classrooms, addressing the challenges of real-time response, issue prioritization, and large-scale personalization.
\end{itemize}

\section{\MakeUppercase{related work}}
Our work intersects with and draws inspiration primarily from three areas: personalized feedback at scale, learning analytics in synchronous classrooms, and instructors' roles in classroom activities. In this section, we review the relevant literature and identify the key needs and challenges in these domains.

\subsection{Personalized Feedback at Scale}

Providing timely and effective feedback is crucial for enhancing student learning. Feedback aims to ``reduce discrepancies between current understandings/performance and a desired goal'' and has one of the most powerful influences on student learning~\cite{hattie2007power}. Effective feedback must be accurate, timely, and tailored to students' needs~\cite{chen2020edcode}. Delayed or poor feedback, such as negative or incorrect feedback, can lead to worse performance than no feedback at all.

Creating real-time, personalized feedback is particularly challenging due to the multifaceted data involved.
In our context of group coding exercises, this includes both coding and natural language (e.g., discussion messages) and group dynamics (e.g., different roles). 
In practice, teachers using LA tools typically focus on individual student performance or the entire class as a whole~\cite{wu2024impact}. As a result, their responses are often limited to either individual or whole-class scaffolding, without fully leveraging the data and setting of collaborative coding exercises. This leaves significant potential untapped, as the primary goal of group discussions is to utilize peer interactions to reduce teachers' workload both for understanding the students' issues and provide peer support~\cite{wang2021puzzleme}.
Therefore, there is a clear need to bridge this gap by utilizing the group setting and co-created data to enhance feedback mechanisms effectively.

On the other hand, the recent advances in AI have led to various approaches in feedback creation for code understanding~\cite{nam2024using} and auto feedback generation~\cite{gabbay2024combining}. While promising, studies have shown that the quality and reliability of auto-generated feedback can be as low as 50\% accurate in assessing the correct mistakes students made in programming assignments~\cite{estevez2024evaluation}. Additionally, students often find AI-generated feedback hard to understand, unfriendly, and requiring manual validation~\cite{nguyen2024beginning}. Instructors also worry that fully automated feedback generation could lead to the misuse of the responses and are concerned about the general relevance and alignment with learning objectives.
This indicates a need for instructors to review feedback before sending it out.

\subsection{Learning Analytics in Synchronous Classrooms}

Learning Analytics (LA), a burgeoning field focused on analyzing and visualizing learner data to enhance educational outcomes, offers educators innovative ways to comprehend and improve the learning process~\cite{clow2013overview}.
This discipline has seen significant research interest in the past decade within the HCI community, driven by the substantial increase in available learner data and the emphasis on quantitative management strategies. 
While integrating analytics into existing learning tools and routines can increase access rates to the analytics, but may not guarantee meaningful engagement without better strategies to manage analytic timing~\cite{jung2024probing}.
Our project builds on previous research in real-time learning support and analytics systems, which have been crucial in understanding student performance. 
These systems are generally divided into two categories: individual behavior and collaboration behavior.

\textbf{Individual Behavior Analytics}. Individual student behavior can be divided into two contexts. First, tools like EduSense~\cite{ahuja2019edusense}, AffectiveSpotlight~\cite{murali2021affectivespotlight}, and Glancee~\cite{ma2022glancee} analyze student and audience facial or body gestures to provide new insights to instructors. These studies have shown that providing rich data about student and audience engagement can help instructors and presenters manage the challenges of large audience sizes. Second, when students engage in in-class exercises, analytics tools should help instructors understand their performance based on task-relevant activities. Tools like Lumilo~\cite{holstein2018classroom}, VizProg~\cite{zhang2023vizprog}, and Codeopticon~\cite{guo2015codeopticon} offer live insights into student activities, such as coding and math exercises, allowing instructors to address student issues in real-time and develop effective teaching strategies.

\textbf{Collaboration Behavior Analytics}. The HCI and Education communities have emphasized the need for collaboration analytics systems that are theoretically grounded, adaptively capture comprehensive interaction data, model collaboration context-sensitively, respect users ethically, and provide customized support for individuals and groups~\cite{schneider2021collaboration}. In response, VizGroup~\cite{tang2024vizgroup} analyzes peer instruction data in in-class coding exercises and automatically summarizes and recommends collaboration behavior events for instructors in real-time. Pair-Up~\cite{yang2023pair}, Groupdynamics~\cite{sato2023groupnamics}, and ClassInsights~\cite{ngoon2024classinsights} have explored various aspects of collaboration analytics by aggregating and displaying group activities such as discussions and learning statuses.

While these systems effectively inform instructors about classroom situations, many teachers also experience difficulty in determining what action to take in response. 
As Karademir et al. found, school teachers often lack the time to provide individual feedback to each student due to limited time resources~\cite{karademir2024don}. Consequently, their attention is primarily focused on weaker students, and they rely heavily on their experience to formulate feedback, which is often not scalable, sustainable, or sometimes biased.
This highlights the necessity for tools that enable educators to quickly and reliably develop personalized feedback for students at all levels of expertise.

%\subsection{Learning Analytics in Synchronous Classrooms}

%Learning Analytics (LA), a burgeoning field focused on analyzing and visualizing learner data to enhance educational outcomes, offers educators innovative ways to comprehend and improve the learning process~\cite{clow2013overview}.
%This discipline has seen significant research interest in the past decade within the HCI community, driven by the substantial increase in available learner data and the emphasis on quantitative management strategies. 
%While integrating analytics into existing learning tools and routines can increase access rates to the analytics, but may not guarantee meaningful engagement without better strategies to manage analytic timing~\cite{jung2024probing}.
%Our project builds on previous research in real-time learning support and analytics systems, which have been crucial in understanding student performance. 
%These systems are generally divided into two categories: individual behavior and collaboration behavior.

\subsection{Instructors' Role and Effort in Classroom Activities}

Human instructors play an irreplaceable role in educational settings, particularly in in-class activities~\cite{comas2011learning, ccardak2016increasing}. Unlike AI tools, instructors possess a unique ability to understand and adapt to the complex cognitive and emotional landscapes of their students, tailoring feedback based on each student's progress and challenges. However, while conducting in-class activities offers more opportunities for teacher-student engagement, research shows that the variety of student behaviors during these activities demands significant effort and flexibility from teachers~\cite{comas2011learning}. This includes effectively monitoring student activities and making appropriate decisions in real-time~\cite{van2010scaffolding}.
There is a pressing need for strategies and tools to support instructors in efficiently identifying and acting upon critical issues during exercises, especially at scale. By addressing this gap, we aim to develop approaches that can augment human instructors' capabilities in real-time, enhancing the quality and effectiveness of interactive learning experiences in programming education while preserving the irreplaceable role of human instructors.

\subsection{Summary}

Based on this prior work, we identified four instructors needs that guided our subsequent formative investigation into the challenges of meeting these needs:

\begin{itemize}
    \item Need 1: real-time analysis of collaborative coding activities to identify critical issues across multiple students.
    \item Need 2: efficient instructor review and customization of AI-generated feedback to ensure quality and alignment with learning objectives.
    \item Need 3: scalable, personalized feedback generation in live, in-class coding exercises.
    \item Need 4: tools that help instructors prioritize their attention on the most critical issues in large programming classes.
\end{itemize}
\section{\MakeUppercase{formative study}}

Building upon the needs identified in prior research, we conducted a targeted formative study to contextualize these challenges within the specific setting of in-class collaborative programming exercises at the university level. This study aimed to bridge the gap between general educational research and the unique demands of real-time, collaborative coding environments.
We engaged four experienced instructors of introductory programming courses at universities in in-depth interviews lasting 30 to 60 minutes each. Our discussions focused on three key areas: 1) The instructors' strategies for managing in-class programming exercises, particularly in light of the real-time feedback challenges highlighted in the literature, 2) their criteria for identifying critical issues during these sessions, expanding on the need for efficient issue prioritization in large classes, 3) specific obstacles they face when providing personalized feedback to struggling students, especially in the context of collaborative coding activities.

\subsection{Instructors want to understand student's mastery of programming concepts and identify learning challenges}

When conducting peer programming sessions, instructors expressed a strong interest in understanding the extent to which students are mastering the material. They are particularly focused on identifying topics that students have mastered versus those that are causing confusion. One of their main goals is to \textit{``understand what students are understanding and what they're not understanding (P1)''}. They also want to see evidence that highlights these struggles, such as error messages and patterns in those errors. As one instructor mentioned, they are interested in \textit{``whether there is some sort of connection between... groups of error messages... and how many people are having that type of error (P2)''}. Other critical concerns include inactivity in peer interactions, lack of engagement with exercises, and instances of code copying. Some instructors are also keen on tracking a student group's progress relative to others. For example, one instructor noted, \textit{``When I see one group doing really well relative to another group, or another group really struggling compared to the rest of the class, that helps me decide when to step in or take action (P3)''}. These insights highlight the need for an organized system that allows instructors to easily identify and address critical issues among students.

\subsection{Instructors finds it challenging to focus on large number of students at the same time}

In-class programming exercises are often conducted in lectures with a large number of students, making it challenging for instructors to understand each student's individual struggles. Instructors reported feeling overwhelmed by the sheer volume of data they need to analyze to gauge each student's progress. One instructor expressed, \textit{``I feel like since I have so many groups, there's no way I'm doing justice to getting a good representation of the various issues that might be on the table (P1)''}. This difficulty extends to tracking student conversations and identifying specific coding issues they encounter. Another instructor noted that the limited timeframe of in-class programming sessions further complicates the analysis of student struggles: \textit{``There just is not sufficient time to fully integrate or take advantage of what you could have learned from the student's answers (P4)''}. These challenges highlight the need for tools or methods to help instructors efficiently understand and manage student progress without feeling overwhelmed. This insight informs DG1, where we want to guide instructor's attention in noticing critical issues in programming classrooms. 

\subsection{Instructors' Strategies for Personalized Guidance and Support in Programming Exercises}

Instructors use a variety of feedback and intervention strategies to ensure students gain the most from programming exercises. One common approach is providing guidance on specific implementation issues. For example, one instructor shared, \textit{``You could guide them to check their spelling or see if they have named a variable incorrectly or used two different variables (P2)''}. Another strategy involves regrouping students to enhance peer interactions and support. As described by an instructor, they might \textit{``take those people aside and possibly regroup them or talk to them, or even pair them with an instructor or expert who would be a good fit (P3)''}. In general, instructors prefer to take action after identifying struggling students, aiming to provide targeted assistance. Some instructors seek a more structured approach to intervene and help those who need additional support. One emphasized the need for clear guidelines for teaching assistants: \textit{``I try to provide very clear instructions for TAs so that all of them hear the same message about how they should handle situations involving struggling students (P4)''}. This underscores the importance instructors place on delivering personalized feedback and support tailored to each student's unique needs. This informs our DG2 and DG3 where instructors should have control over the support and feedback they give to students based on the student's performance, and during the process they should be given ways to align their intention to the feedback.

\section{\MakeUppercase{Design Goals}}

Based on our formative study with instructors and the identified gaps in existing literature, we established the following design goals to guide the development of \sys{}:

\begin{itemize}
    \item \textbf{DG1: Effectively Guide Instructor Attention to Critical Issues in Real-Time.} To address the challenge of information overload in large-scale programming sessions, the system should help direct instructors' attention to the most significant issues and patterns emerging from students' collaborative activities with minimal effort. This goal aims to enable instructors to maintain cognitive control while leveraging system support to navigate complex, real-time programming environments.
    
    \item \textbf{DG2: Empower Instructors with Guided Control over LLM-Generated Feedback.} To bridge the gap between LLM capabilities and effective personalized feedback, the system should offer intuitive mechanisms for instructors to guide and control the feedback generation process. This goal focuses on enabling instructors to shape AI-generated feedback based on key pedagogical components, ensuring alignment with their instructional intentions and the specific context of the programming session.
    
    \item \textbf{DG3: Reduce the Time and Effort Required to Ensure the Quality of LLM-generated Feedback.} To mitigate risks associated with AI-generated content, such as hallucinations or misalignment with learning objectives, the system must facilitate quick identification and rectification of issues in AI-generated feedback. This goal aims to provide efficient methods for instructors to verify the accuracy, relevance, and pedagogical alignment of feedback before it reaches students, ensuring quality without sacrificing the scalability benefits of AI assistance.
\end{itemize}

\begin{figure*}
    \centering
    \includegraphics[width=\textwidth]{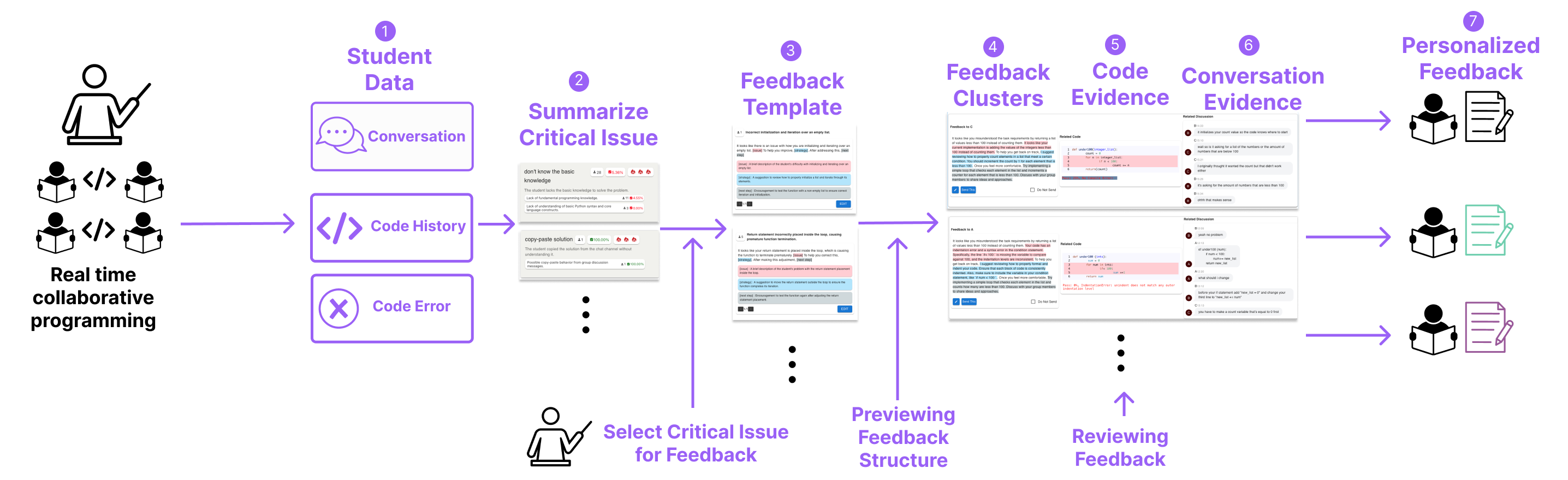}
    \caption{\sys{}'s Workflow Overview. Once students' conversation logs, code history, and code errors come in (1), \sys{} continuously identifies critical issues and recommends them to the instructors (2). Instructors select the critical issues for feedback, which are then summarized and categorized to create Feedback Templates (3). These templates are previewed by instructors and further clustered (4) with relevant Code Evidence (5) and Conversation Evidence (6) to provide context and support a rapid verification process. This results in personalized feedback being sent to each student (7). }
    \label{fig:sys_workflow}
\end{figure*}

\section{\MakeUppercase{system}}
Based on our design goals, we developed \sys{}, a system that guides instructors' attention to critical issues of students' coding and group discussion activities, and facilitates the creation and review of personalized feedback. \sys{}'s UI consists of two main components: a Critical Issue Recommendation Panel (Figure ~\ref{fig:criticalIssue_panel}), and a Structured Feedback Creation and Review Panel (Figure ~\ref{fig:feedback}). We begin this section with an overview of \sys{}'s workflow, and then describe the design and implementation of the system in detail. 

Figure~\ref{fig:sys_workflow} illustrates \sys{}'s system workflow. During a class exercise, \sys{} captures students' code submissions and messages sent in small group chats when group discussion is enabled (1). Simultaneously, \sys{} continuously analyzes this data, classifying students' coding activities and group interactions into predefined issues, ranking their severity, and recommending critical ones to instructors (2). Upon seeing these recommendations, instructors can browse the issues and select one to provide feedback on. 
\sys{} facilitates feedback creation by providing key feedback components so that instructors can easily select multiple of them to create a LLM prompt (3).
This prompt is sent to a custom LLM along with its associated context, which returns a list of personalized feedback suggestions.
\sys{} clusters similar feedback instances based on their associated issues (5), and presents each cluster and its context in a grid view (4, 5, 6).
Once instructors complete the review process, they can send the finalized, personalized feedback to the relevant student(s) as a chat message(7).

\subsection{Critical Issue Recommendation Panel}
\begin{figure*}[h]
    \centering
    \includegraphics[width=1.05\linewidth]{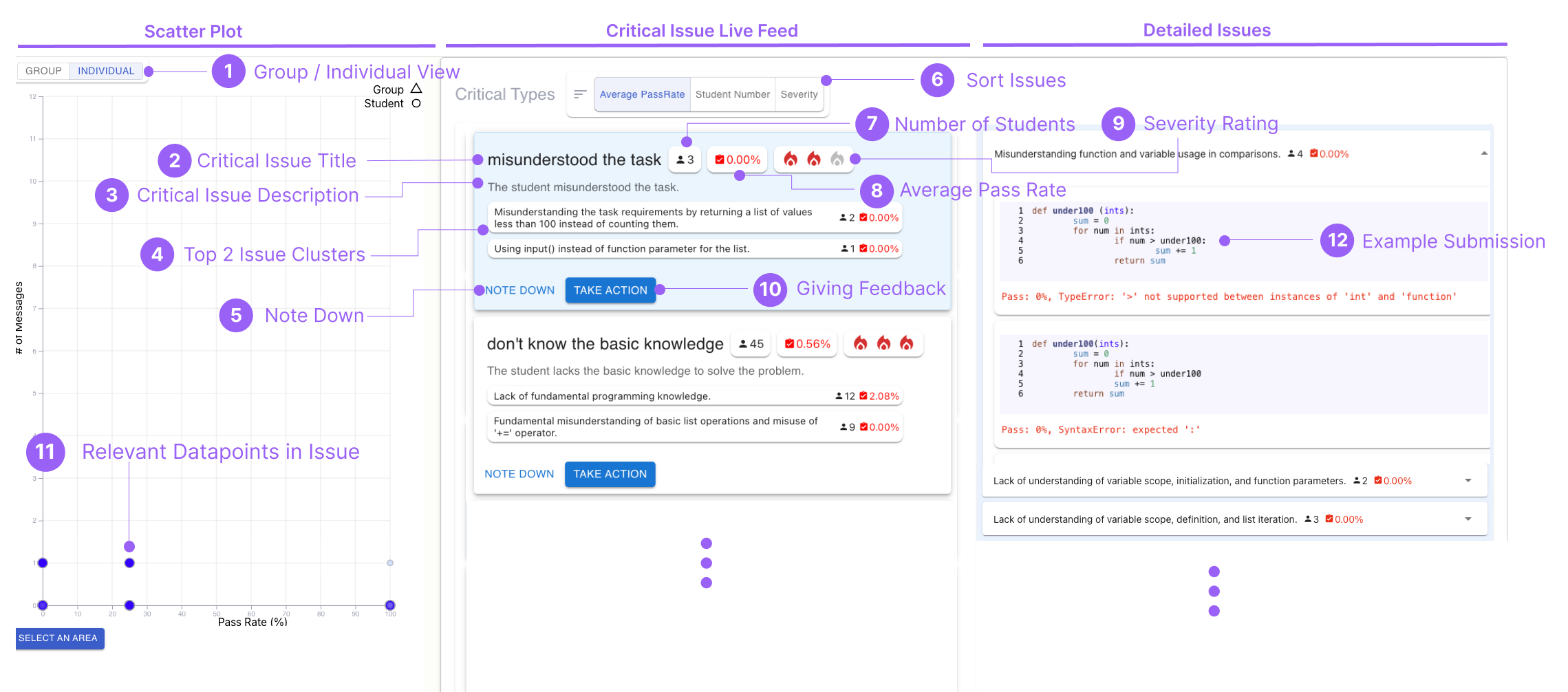}
    \caption{Critical Issue Recommendation Panel consists of Scatterplot, Critical Issue View and Detailed Issue. (1) Scatterplot changes based on group view. Critical Issue component consists of (2) Name of the Critical Issue, (3) Description of the Critical Issue, (4) Top 2 sub issue, (7) Number of Students with this critical issue, (8) Average passrate of students with this critical issue. Instructors can (5) note down and save the critical issue or (9) give feedback to the critical issue. Instructors can sort critical issue based on pass rate, number of students or severity. (10)Selecting a critical issue would display example submission and (11)highlight relevant datapoints. }
    \label{fig:criticalIssue_panel}
\end{figure*}

%Need to define critical
To effectively guide instructor attention to critical issues (DG1), \sys{} employs a custom LLM that continuously identifies patterns in students' coding activities and group interactions, classifies those that are critical, and then displays them on a teacher dashboard (Fig.~\ref{fig:criticalIssue_panel}). 
Teacher dashboards are visual displays that show student learning activities and progress.
There are extensive studies on how to design an effective teacher dashboard and how they can help teachers inform their interventions and decision making~\cite{greller2012translating,van2022teacher,wise2019teaching}.
Building upon prior work, \sys{}'s dashboard contains a high-level scatter plot showing students'/groups' pass rates and the number of messages they have sent (left), a list of curated critical issues detected dynamically by the system (middle), and detailed information for each critical issue upon user click (right). The design of the scatter plot is inspired by prior work that shows students' coding and group discussion progress~\cite{tang2024vizgroup,zhang2023vizprog}, while the design of the list view for live events is adopted from learning analytics literature that iteratively designed for instructors to be aware of students' ongoing events during exercises~\cite{aleven2022dashboard}.

\subsubsection{Activity and Trends (Scatter Plot)}
Current activity and historical trends of individual and group activities are displayed via a scatter plot, which enables instructors to identify collaborative patterns and recognize students who may need additional support. In the Group view, the scatter plot shows the total number of messages versus the average group pass rate. In the Student view, it plots the number of messages sent by each student against their respective pass rates (Figure ~\ref{fig:criticalIssue_panel}.1).

\subsubsection{Critical Issues}
To facilitate easy identification of student struggles and help instructors prioritize their attention in large classes (DG1), critical issues are displayed in a live feed that updates in real time as students work on programming assignments. These critical issues are dynamically summarized using an LLM architecture based on student activities (Section 5.2). Similar to other learning analytics systems, such as VizProg and CFlow\cite{zhang2023vizprog,zhang2024CFlow}, the critical issues are organized to allow instructors to inspect them at varying levels of granularity. This design ensures instructors can gain a high-level understanding of student performance before understanding the low level details of the student's critical issue.

Each entry in the feed includes the name of the critical issue (Figure ~\ref{fig:criticalIssue_panel}.2), a brief description (Figure ~\ref{fig:criticalIssue_panel}.3), the number of students affected (Figure ~\ref{fig:criticalIssue_panel}.7), the average pass rate for the relevant group (Figure ~\ref{fig:criticalIssue_panel}.8) and the severity of the issue (Figure ~\ref{fig:criticalIssue_panel}.9). More detailed issues with lower pass rates or higher numbers of students are aggregated under these broader categories (Figure ~\ref{fig:criticalIssue_panel}.4). Instructors can sort the live feed by average pass rate, number of students, and issue severity (Figure ~\ref{fig:criticalIssue_panel}.6). Clicking on an issue reveals all the clustered sub-issues under it, with the corresponding data points highlighted in blue on the scatter plot (Figure ~\ref{fig:criticalIssue_panel}.11).

Each sub-issue displays the number of students involved and the average pass rate. Expanding these sub-issues reveals the relevant code snippets or student conversation data (Figure ~\ref{fig:criticalIssue_panel}.12). Below each code snippet, the system shows the pass rate and any compilation errors associated with the code. After reviewing the critical issues, instructors have the option to save them for future reference (Figure ~\ref{fig:criticalIssue_panel}.5), or proceed to giving feedback to students with the specific issue (Figure ~\ref{fig:criticalIssue_panel}.10).

\subsection{Critical Issue Recommendation Model}

Prior work has shown LLMs' potential to identify and classify students' challenges~\cite{10.1145/3636555.3636905}. To ensure that the LLM can effectively identify students with critical issues, we annotated two previously recorded sessions of students completing a programming exercise. This data is collected from two live collaborative programming sessions conducted at a local university. The collected data includes student's coding activities, and conversation history. Two members of our research team annotated the recorded student activity data using an annotation tool (Appendix ~\ref{sec:annotator A}). This tool allows annotators to watch a playback of different segments of student code and messaging history, and label student activity. We will make this tool available for public use. 

\textit{Dataset: }
Both sessions involved over 70 students, with more than 30 minutes of student activity data per session. Student activity was automatically segmented during the recording session. The two session combined result in over 502 data points. 

\textit{Annotation Process:}
For each segment, annotators were first asked to read the task description and identify the learning objectives of the task, such as function use, if/else statements, loops, variable initialization, etc. Then, for each student, annotators examined each segmented activity and watched the replay. They considered contextual factors such as code, discussion, error messages, pass rates, and overall class performance before determining whether the situation was critical and providing a label for the critical situation.We defined a critical situation as one where instructors' attention is needed and students can benefit from the instructor's help. For each identified critical situation, we asked annotators to identify the related code issues or conversation issues. The different types of issues are shown in Table ~\ref{table:issuetaxonomy}. These issues are determined based on prior literature of challenges student's face during collaborative sessions, and the formative study we conducted on instructors regarding critical issues they are interested in during in-class programming sessions. Annotators were instructed to consider the alignment of the issues with the activity type (e.g., message labels should be related to the student's group discussion). 

The two annotators first conducted rounds of initial coding independently (including pilot coding sessions), and met multiple times to discuss their annotations for clarifications and to resolve any discrepancies. We then performed inter-rater reliability test for the 103 initially coded student activity data points which cover 20\% in overlapped coding. Out of the 103 initially coded data,  we report a Cohen-Kappa score for determining whether an issue is critical or not of 0.85. Across all labels (detailed in Table ~\ref{table:issuetaxonomy}), we report a mean Cohen-Kappa score of 0.78 ($S.E. = 0.076$). Based on the minimum Cohen-Kappa score across all label (0.53 for Syntax Error label), the annotators reached a moderate agreement and thus the annotators proceeded with independent coding for the remaining 400 labels.

\textit{Data Training: }
After the labels were finalized, we randomly split the dataset into training and test sets, containing 80\%, 20\% of the original dataset, respectively. We build on this pipeline by incorporating human annotators to improve the quality of the training data so that it will better align with instructor’s intent. The training data are used as few-shot examples of a prompt that categorize student issues (Appendix B). We used GPT-4o in the data training process, and state-of-the-art prompting engineering techniques, such as few-shot prompts~\cite{NEURIPS2020_1457c0d6} and AI-chains~\cite{10.1145/3491102.3517582}, were adopted. After evaluating the model with test data, we report an accuracy score of 0.71 for the LLM identifying issues that are critical and an $F1_{weighted}$ score of 0.57.

\subsection{Personalized Feedback Creation \& Review Panel}
\begin{figure*}[h]
    \centering
    \includegraphics[width=1\linewidth]{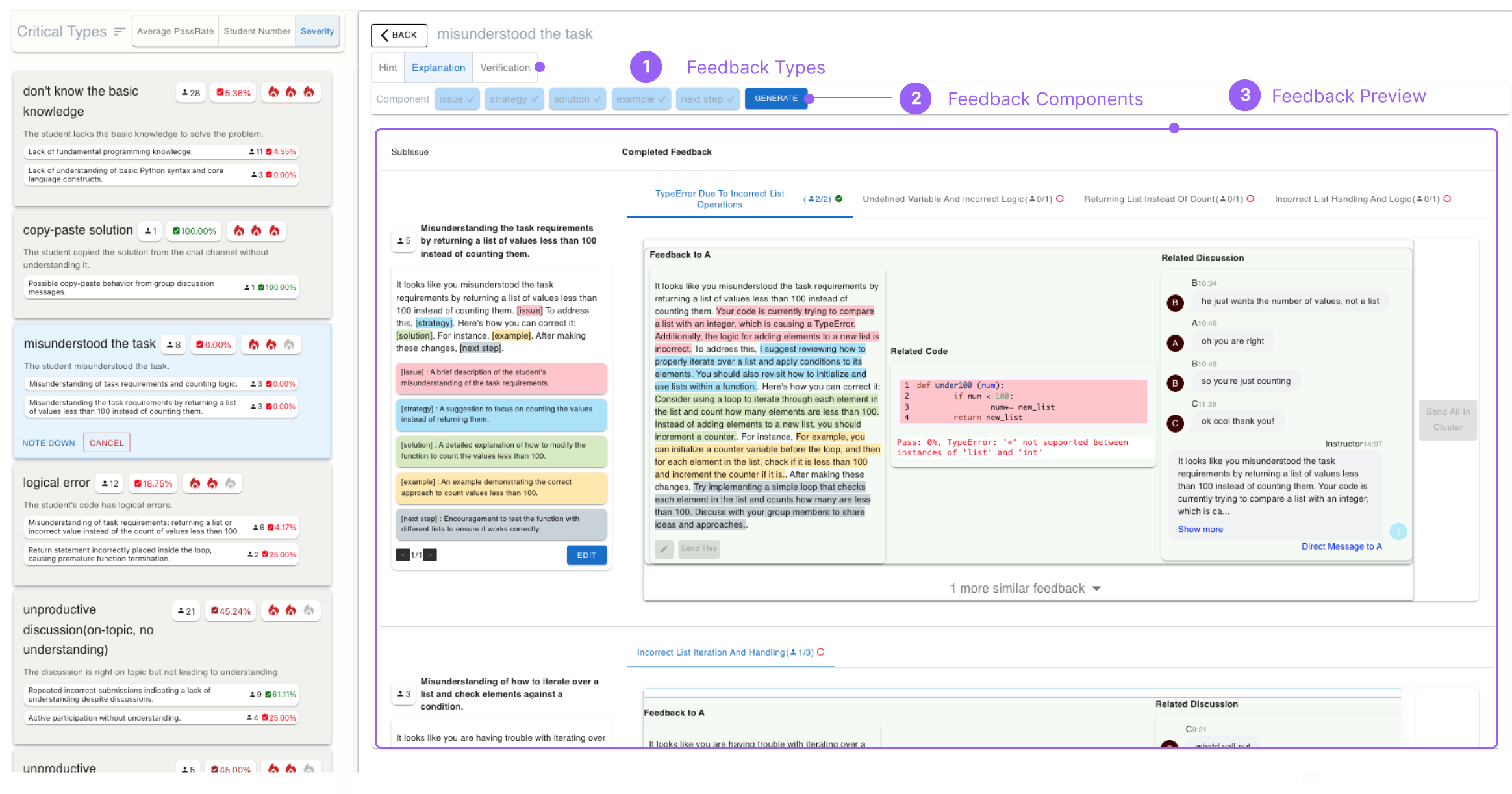}
    \caption{User Interface for creating and reviewing LLM-generated feedback. (1) Instructor can select different feedback types, which will select a subset of (2)feedback components. (3)Instructors can then generate feedback for preview.}
    \label{fig:feedback}
\end{figure*}
When selecting an action, instructors can provide feedback on the currently selected issue. To facilitate personalized feedback generation (DG2), we developed a `strategy-detail-verify' workflow. First, instructors can strategize by choosing the most appropriate feedback mode from three options: Hints, Explanation, or Verification (Figure ~\ref{fig:feedback}.1). Upon pressing the generate button, a template for the specific feedback type is created, personalized for each sub issue, with placeholders inserted at designated positions. Next, feedback is generated in smaller clusters for each subissue, allowing instructors to see the details of the feedback intended for each student. Related code examples and conversations are displayed alongside the feedback. Finally, instructors can verify the validity of the feedback before deciding whether to send it to students. This step ensures the appropriateness and accuracy of the feedback.

\subsubsection{Component Based Feedback Generation}
\begin{table*}[htp]
\caption{Feedback component description for the issue: \textbf{Lack of understanding of basic programming concepts and Python syntax.}}
\label{table:feedback_taxonomy}
\Description{A taxonomy classifying feedback on programming concepts and Python syntax, including categories for Issue, Strategy, Solution, Example, and Next Step.}
\begin{tabular}{lll}
\hline
\textbf{Component} &
  \textbf{Definition} &
  \textbf{Example Feedback} \\ \hline
\colorbox[HTML]{fec6cc}{Issue} &
  \begin{tabular}[c]{@{}l@{}}A brief description of the student's difficulty\\ with basic programming concepts and Python\\ syntax.\end{tabular} &
  \begin{tabular}[c]{@{}l@{}}Your code is not correctly counting the numbers less\\ than 100. Specifically, you are using the $<=$ operator\\ instead of $<$ to increment the count, which is causing\\ the logic to fail.\end{tabular} \\ \hline
\colorbox[HTML]{abe2fa}{Strategy} &
  \begin{tabular}[c]{@{}l@{}}A suggestion to focus on understanding\\ fundamental programming principles and\\ Python syntax.\end{tabular} &
  \begin{tabular}[c]{@{}l@{}}To address this, I recommend focusing on how to\\ properly iterate through the list and maintain a separate\\ counter variable to keep track of how many numbers\\ are less than 100.\end{tabular} \\ \hline
\colorbox[HTML]{d8eac3}{Solution} &
  \begin{tabular}[c]{@{}l@{}}An example of how to define a function and\\ use loops in Python.\end{tabular} &
  \begin{tabular}[c]{@{}l@{}}You can create a counter variable initialized to zero\\ and increment it each time you find a number less\\ than 100. This way, you will have the correct count to\\ return at the end of your function.\end{tabular} \\ \hline
\colorbox[HTML]{ffe8ae}{Example} &
  An Example implementation of the solution &
  \begin{tabular}[c]{@{}l@{}}For instance, if you have a list like [50, 150, 75, 200],\\ using the correct list comprehension will yield [50, 75],\\ and then you can use \texttt{len()} to count these elements.\end{tabular} \\ \hline
\colorbox[HTML]{c9d3d7}{Next step} &
  \begin{tabular}[c]{@{}l@{}}Encouragement to implement the 'under100'\\ function after gaining confidence with the\\ concepts.\end{tabular} &
  \begin{tabular}[c]{@{}l@{}}Try implementing the corrected version of your\\ function and test it with different lists of integers to\\ ensure it works as expected. This will help reinforce\\ your understanding of the counting logic.\end{tabular} \\ \hline
\end{tabular}
\end{table*}
Feedback is structured to consist of up to 5 components: \colorbox[HTML]{fec6cc}{Issue}, \colorbox[HTML]{abe2fa}{Strategy}, \colorbox[HTML]{d8eac3}{Solution}, \colorbox[HTML]{ffe8ae}{Example}, \colorbox[HTML]{c9d3d7}{Next step}. The definition of each component is detailed in Table \ref{table:feedback_taxonomy}. These five components are derived from Hattie et al.'s feedback model ~\cite{hattie2007power}, which states that effective feedback answers three questions: \textit{(1) What are the goals? (2) What has been done to progress towards the goals? and (3) What activities need to be undertaken to make better progress?}

To support instructors in understanding what students have done to progress through the exercise, we included the \colorbox[HTML]{fec6cc}{Issue} component in the feedback, which summarizes students' current problems in their code. Example issues include student misunderstanding of important programming concepts such as for-loops or disengagement in group discussions.

To ensure that the feedback provides actionable steps for students to follow, which is derived from the question 'What activities need to be undertaken to make better progress?', the feedback template contains a \colorbox[HTML]{abe2fa}{Strategy} component. This component offers suggestions on how to resolve the current issues students are facing. \colorbox[HTML]{abe2fa}{Strategy} also highlights the goal the student is trying to achieve, addressing the question 'What are the goals?' from the feedback model.
\colorbox[HTML]{d8eac3}{Solution} and \colorbox[HTML]{ffe8ae}{Example} build upon \colorbox[HTML]{abe2fa}{Strategy} by providing students with additional examples to solidify their understanding. This approach aligns with prior research showing that code examples and walkthroughs of similar problems are effective in improving students' understanding ~\cite{Robins2003LearningAT}. Prior feedback systems have emphasized the importance of positive feedback in increasing intrinsic motivation ~\cite{karademir2024don}. Hence, the \colorbox[HTML]{c9d3d7}{Next step} component in the feedback provides positive encouragement and affirmation to the student in incorporating the feedback.

Furthermore, each piece of feedback is directed at three levels respectively: Task Level, Process Level, and Self-Regulation Level ~\cite{hattie2007power}. The Task Level refers to how well the tasks are understood, the Process Level refers to the process needed to understand and perform the task, and the Self-Regulation Level refers to self-monitoring, directing, and regulating actions. Similar to prior work on feedback systems in learning analytics dashboards, these levels are taken into account during feedback generation.
Instructors often give feedback to guide students who are stuck, clarify students' confusion, or check students' understanding. Hence, depending on each student's situation, instructors can provide three types of feedback: Hints, Explanations, or Verifications.

\begin{enumerate}
    \item \textbf{Hints: }Hints guide students who are stuck in solving related programming tasks by pointing out their current issues and providing next steps.
    \item \textbf{Explanation: }Explanation-type feedback first identifies the issue the student is facing, then provides a detailed explanation of how to approach the problem, and gives students concrete next steps.
    \item \textbf{Verification: }The Verification component encourages students to solve a similar problem to solidify their understanding of the task objectives.
\end{enumerate}
Table 2 includes example feedback for each type. Providing such feedback presets allows instructors to quickly generate feedback that fits their teaching strategy, thereby reducing the time required to draft feedback (DG3).

\subsubsection{Feedback Generation Pipeline}
\begin{figure*}[h]
    \centering
    \includegraphics[width=1\textwidth]{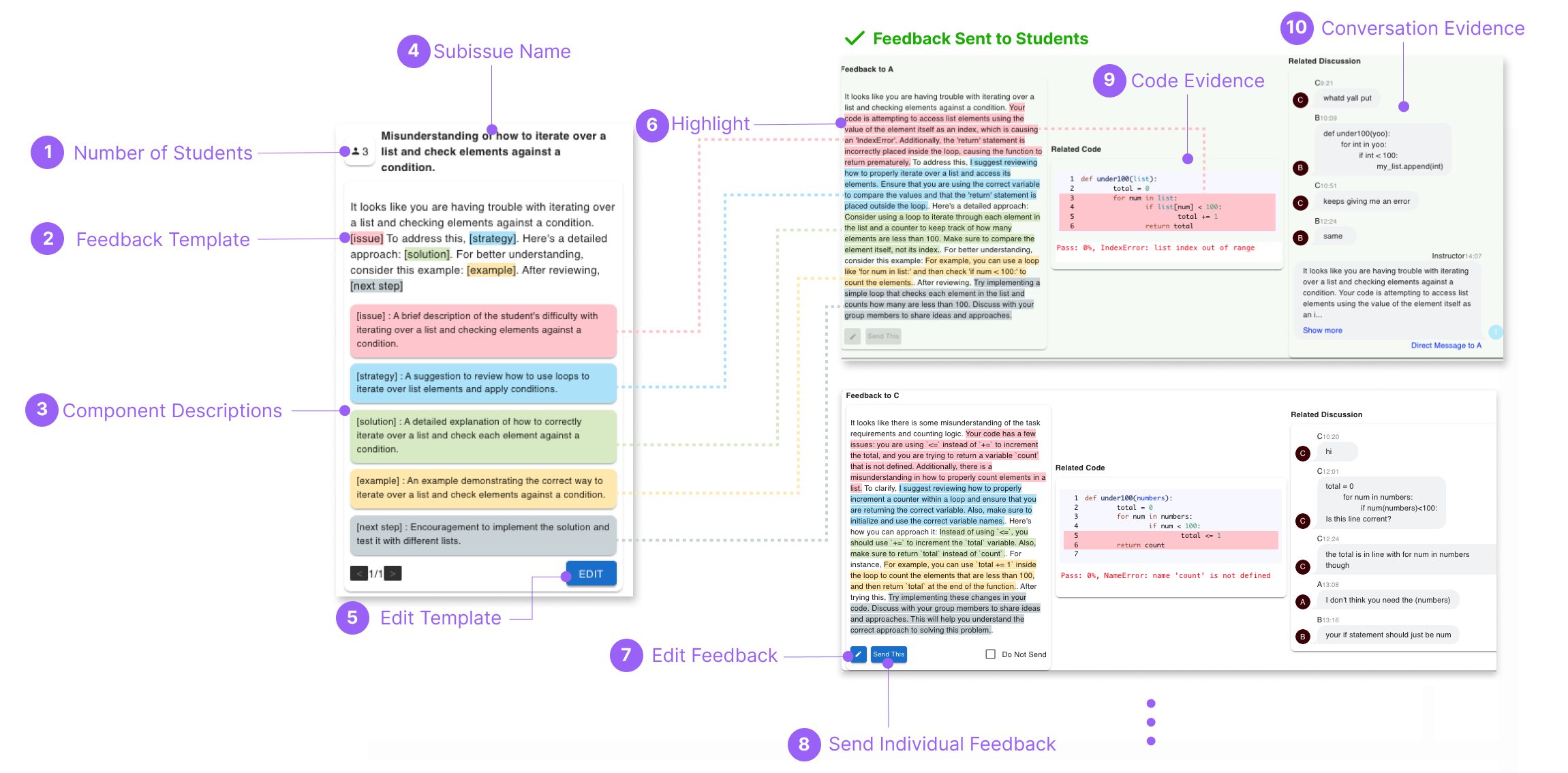}
    \caption{Feedbacks are generated for each sub issue, which includes the name of the sub issue (4), the number of students in the sub issue(1), the feedback template with components as placeholders (2) and component description (3). Instructors are able to edit the template (5). Feedback are then generated in clusters that includes code evidence (9) and conversation evidence (10). Feedback for each components are highlighted with their corresponding colors (6). Instructors can choose to edit or send individual feedback (7, 8).}
    \label{fig:criticalIssue}
\end{figure*}
Feedback is generated using two LLM prompts: one for generating each component of the feedback and another for combining the generated components into the final feedback (Appendix ~\ref{appendixE}) using few-shot prompting. Both prompts include guidelines for generating feedback derived from the feedback model. We used the GPT-4 model in generating such feedback. For clustering, we utilized k-means clustering with the text-embedding-3-large model, employing 100 iterations. The feedback generation process is as follows:

\textit{Step 1: Feedback Template Generation:  }
\sys{} takes in the issue description, feedback type and chosen components, and generates a feedback template with selected content as placeholders (in [square brackets]), as well as feedback corresponding to each component. For instance, with the issue description of \textit{Lack of understanding of basic list operations and iteration}, feedback type of \textit{Hint}, and chosen component of \textit{\colorbox[HTML]{fec6cc}{Issue}, \colorbox[HTML]{abe2fa}{Strategy} and \colorbox[HTML]{c9d3d7}{Next step}}, \sys{} would generate a feedback template such as \textit{"It seems you are having some trouble with basic list operations and iteration. \colorbox[HTML]{fec6cc}{[Issue]} To help you get back on track, \colorbox[HTML]{abe2fa}{[Strategy]}. Once you feel more comfortable, \colorbox[HTML]{c9d3d7}{[Next step]}"}. And individual feedback on each component, 
\begin{enumerate}
    \item \textbf{\colorbox[HTML]{fec6cc}{Issues}: }A brief description of the student's difficulty with basic list operations and iteration,
    \item \textbf{\colorbox[HTML]{abe2fa}{Strategy}: }A suggestion to review list iteration and applying conditions to list elements.
    \item \textbf{\colorbox[HTML]{c9d3d7}{Next step}: }Encouragement to implement a simple loop to check each element in the list. 
\end{enumerate}

\textit{Step 2: Detailed Feedback Composition}
\sys{} then takes the generated list of component feedback and apply it to the template, generating a complete feedback for students. It also takes in student information and elaborates on student’s performance by referencing their coding and conversation history. For instance, the issue the use of \texttt{input()} and \texttt{split()} is not necessary for this task will lead to highlighting of the related lines with functions \texttt{input()} and \texttt{split()}. 

\subsubsection{Feedback Clusters with highlights on Evidence}
%Add citations ono evidence for AI
Our design for displaying feedback clusters aligns with prior literature on designing UI elements for validating large-scale LLM outputs, which allows users to determine whether the AI-generated feedback meets their current needs ~\cite{glassman2024airesilientinterfaces}. Feedback is generated in clusters to enable instructors to easily review the generated content (DG3).
Under each sub-issue within a critical issue (Figure ~\ref{fig:criticalIssue_panel}.12), feedback is further clustered into smaller groups to ensure that instructors can easily review different feedback clusters. These clusters are created using k-means embedding for semantically similar issues.
To facilitate easy verification of feedback (DG3), we incorporated highlighting as a way for instructors to review how each component is incorporated into the final feedback, as well as the related code or conversation that reflects the feedback. This design is inspired by prior work that supports sensemaking in large amounts of LLM outputs, which is similar to our current use case of enabling understanding of feedback variations given to a cluster of students ~\cite{Gero2024Sensemaking}.

\begin{table*}[t!]
    \centering
    \small
    \begin{tabular}{l  c c c  c c c c c c c c c}
    \toprule
     \multirow{2}{*}{\textbf{Condition}}& \multicolumn{3}{c}{\textbf{Generated}}& & \multicolumn{3}{c}{\textbf{Sent}} & & \multicolumn{4}{c}{\textbf{Edited}$^{**}$}\\
     \cline{2-4} \cline{6-8} \cline {10-13}
      & -1 &0 & 1 &   & -1 &0  &1  & &low-low &low-high &high-low &high-high\\
      
    \midrule
    \textbf{Baseline} & $44.17\%$ & $10.00\%$  & $45.83\%$ & &  $45.00\%$&$8.67\%$  & $46.33\%$ & & $45.00\%$&   $20.00\%$ &  $5.00\%$&  $30.00\%$\\
    \textbf{\sys{}} & $14.16\%^{*}$& $5.00\%$ & $80.83\%^{**}$ & & $9.17\%^{**}$&$10.67\%$ & $80.17\%^{**}$ & & $0.00\%$&   $22.22\%$&  $0.00\%$&  $77.78\%$\\
     \bottomrule
    \end{tabular}
    \caption{Feedback Quality in the Sampled Dataset. -1, 0, and 1 denote \textit{incorrect feedback}, \textit{shallow feedback}, and \textit{high-quality feedback} respectively. Edited feedback messages are grouped by the change of qualities before and after the edit, where low-quality feedback includes both \textit{incorrect feedback} and \textit{shallow feedback}. Data of edited changes are tested by the Fisher Exact Probability Test ($p = 0.001$).  $*$ indicates $p < 0.05$, while $**$ indicates $p < 0.01$.}
    \label{tbl:feedback_quality}
\end{table*}
\section{\MakeUppercase{Lab study with instructors}}
We conducted an in-person between-subject user study to examine \sys{}'s usability and effectiveness for identifying students' issues and sending feedback.
Our study is approved by the IRB at our institute.

\subsection{Participants}

We recruited 20 participants (7 female, 11 male, 2 non-binary) who have both teaching and programming experiences at a four-year university via personal networks, local mailing lists, and snowball sampling. During the study, participants were asked to inspect and give feedback to student behavior that was collected in a large programming course at our institution. Each participant was compensated with \$15 USD for their time and effort.

\subsection{Protocol}
\subsubsection{Live Simulation}
To simulate the real-time, in-class teaching experience of a large lecture, we asked participants to watch a live playback of an in-class exercise session. To ensure the authenticity of the data participants interacted with, we used a research tool~\cite{tang2023vizpi} to capture real data from a large introductory-level university programming course's collaborative exercise session that involved 111 students. During the exercise session, students first worked on a Python programming exercise, and then they were divided into groups to discuss their issues and help each other on their laptops. The Python problem was to write a function to count the number of elements less than 100 in a given list.

\subsubsection{Conditions}
We used a between-subject study design where each participant used the system under one of the following conditions:
\begin{itemize}
    \item \textbf{Baseline:} a baseline version of \sys{} without any feedback type, feedback component, and feedback template for generation, and participants control the feedback generation by editing the prompt to instruct the LLM, which is one of the most common ways to use LLM. A default prompt was given to ask AI to generate feedback based on the student's recent code submissions and discussions. There is no visual augmentation for feedback review either, but the baseline version still had the complete Critical Issue Recommendation Panel.
    \item \textbf{\sys{}:} a full version of \sys{} with all its features.
\end{itemize}

\subsubsection{Tasks}
Participants complete three tasks in the study:
\begin{itemize}
    \item \textbf{T1 and T2:} Task 1 (T1) and Task 2 (T2) asked participants to help as many struggling students as they could by checking the critical issue recommendation panel and sending feedback. T1 and T2 were limited to 6 minutes each.
    \item  \textbf{T3:} Task 3 (T3) asked participants to help students who were actively discussing but with low performance by selecting the plot and sending feedback on a certain time stamp.
\end{itemize}
T1 covers the individual programming stage of the exercise session, while T2 and T3 focus on the collaborative programming stage. For all three tasks, we emphasized the importance of the feedback quality to participants and had them understand the stake level of the tasks by asking them to imagine this is the live class setting. T1 and T2 were designed to mimic the real-world circumstances that instructors need to address students' issues while facing the dynamically changed issue set and limited time. Similarly, T3 was designed to simulate a more focused issue resolving process while still need to provide timely support to students.

\subsubsection{Study Procedure}
Each study was conducted in person in a lab setting and lasted around 45 minutes. At the beginning of each study session, we collected informed consent from the participants after introducing the goal and the process of the study. After that, we gave participants an explanation of the context of the data and tasks used in the study. Following the general introduction and explanation, we then offered a detailed tutorial of the system and asked the participant to explore the system on a trial dataset to warm them up and answer their questions. Once they were familiar with the system, participants were asked to complete the three designed tasks. At the end of the study, participants completed a survey with Likert scale questions and participated in a semi-structured interview. All studies were screen- and audio-recorded and participants were asked to think aloud while completing the survey questions. The order of conditions and tasks were counter-balanced and randomized.

\begin{table*}[t]
\centering
\small
\begin{tabular}{l c c c c c c c c}
\toprule
\textbf{Condition} & \textbf{Mental demand} & \textbf{Physical demand} & \textbf{Temporal demand} & \textbf{Performance} & \textbf{Effort} & \textbf{Frustration} \\ 
\midrule
\textbf{Baseline}   & 5.0 (4.60 ± 1.26)    & 2.0 (2.10 ± 1.20)       & 5.0 (4.60 ± 1.17)       & 4.0 (3.90 ± 1.37)    & 4.0 (4.00 ± 1.25)    & 2.5 (2.70 ± 1.57)    \\ 
\textbf{\sys{}} & 5.5 (4.80 ± 1.93)     & 2.0 (2.30 ± 1.49)       & 5.0 (4.60 ± 1.78)       & 3.0 (3.10 ± 1.10)    & 4.0 (4.20 ± 1.48)    & 2.5 (3.00 ± 1.76)    \\
\bottomrule
\end{tabular}
\caption{Response to NASA TLX items. Format: median (mean ± standard deviation)}
\label{tbl:nasa-tlx}
\end{table*}
\subsection{Results}

By recording the feedback participants generated, edited, and sent in the user study, we collected 5871 feedback messages in total (5049 generated, 107 edited, 715 sent). To evaluate the quality of the feedback, two researchers used a custom annotation tool to annotate the feedback data (Appendix~\ref{appendixB}). The annotation process was as follows: First, we sampled feedback data from each participant and each task proportionally ($1\%$ of generated feedback, $20\%$ of edited feedback, and $10\%$ of sent feedback) and randomly shuffled the sampled data. Then, each sampled data point was coded into one of the following three categories: \textit{incorrect feedback (-1)}, \textit{shallow feedback (0)}, and \textit{high-quality feedback (1)}. A description of the coding standards is provided in Appendix~\ref{appendixG}. After initializing the coding scheme, the two researchers first conducted a pilot coding meeting and discussion to clarify and resolve conflicts on the coding standards. Then, they began independently coding the same sampled dataset. For the independent annotations, the agreement on \textit{incorrect feedback}, \textit{shallow feedback}, and \textit{high-quality feedback} was $85.96\%$, $53.33\%$, and $96.15\%$, respectively. To achieve a high level of consensus\cite{10.1145/3359174}, the two researchers then negotiated and resolved each disagreement through discussion until 100\% agreement was reached on the coded sample dataset (Table\ref{tbl:feedback_quality}).

\subsubsection{\sys{} helped participants create better feedback}
Participants using \sys{} sent significantly more \textit{high-quality feedback} (\sys{}: $\mu = 80.17\%, \sigma = 0.17$; baseline: $\mu = 46.33\%, \sigma=0.28$; $p < 0.01$), while they also sent less \textit{incorrect feedback} (\sys{}: $\mu = 9.1717\%, \sigma=0.15$; baseline: $\mu = 45.00\%, \sigma=0.31$; $p < 0.01$) in the sampled sent dataset. 
We did not find evidence of a significant difference in the number of \textit{shallow feedback} between two conditions (\sys{}: $\mu = 10.67\%, \sigma=0.17$; baseline: $\mu = 8.67\%, \sigma=0.14$).

Quantity-wise, we did not observe a significant difference in the number of feedback sent between the two conditions (\sys{}: $\mu = 32.20, \sigma = 12.43$; baseline: $\mu = 24.20, \sigma = 24.93$).

In the open-ended task (T3), the number of sent feedback was on par between two conditions (\sys{}: $\mu = 1.57, \sigma = 0.71$; baseline: $\mu = 1.30, \sigma = 0.59$).

\subsubsection{Control feedback generation}
Based on the sampled generated dataset, under \sys{} condition, participants generated significantly more (\sys{}: $\mu = 80.83\%, \sigma = 0.22$; baseline: $\mu = 45.83\%, \sigma=0.28$; $p < 0.01$) \textit{high-quality feedback} than the baseline while generated less (\sys{}: $\mu = 14.16\%, \sigma = 0.15$; baseline: $\mu = 44.17\%, \sigma=0.32$; $p < 0.05$) \textit{incorrect feedback}. On average, there are  $5.00\%$ ($\sigma=0.11$) \textit{shallow feedback} in feedback generated by \sys{} and $10.00\%$ ($\sigma=0.16$) \textit{shallow feedback} in feedback generated by the baseline.

During the study, participants using \sys{} attempted 2.40 ($\sigma=2.07$) modifying feedback type and component to generate new feedback, while participants using the baseline attempted 1.8 ($\sigma=2.30$) editing the prompts to control the generation results. The survey results show that participants reported high-level perceived controls of the feedback generation process for both \sys{} ($Median = 6.00, \sigma=1.71$) and the baseline system ($Median = 5.50, \sigma=1.76$), as well as the perceived ease of control level (\sys{}: $Median = 5.50, \sigma=1.17$; Baseline: $Median = 6.00, \sigma=1.17$). 

Though we could not obtain a significant difference in the number of attempts to control the generation process, participants' behaviors reflect the difference in their intentions. Under the baseline condition, participants' intentions of controlling generation are verbally expressed in the modified prompt, e.g., \textit{``Try and shorten the response'', ``Don't give away answers''}. The goals are limited to low-level requirements about the length of the generation and not to include the direct solution in the result (P8). Whereas, using \sys{}, participants tended to think about the type of help they would like to offer to students, which is unseen in the baseline condition. For instance, P9 included example generation in the feedback for students who "lacked basic knowledge" in order to help them get started with the problem. P5 suggests that they would consider excluding some components (i.e., Next Steps) in the feedback depending on the severity of the issue. Likewise, P3 continued reflecting on the feedback strategy while completing the survey and talked about \textit{``oh, maybe I should use verification in that case (to give feedback about cannot correct by just saying words)''}. Moreover, participants also commented on the feedback \sys{} generated as like `\textit{`from a senior instructor''}(P7).

\subsubsection{Changes in feedback edits}
During the user study, participants made $6.20$ ($\sigma=8.75$) edits using \sys{}, and participants using the baseline edited $4.50$ ($\sigma=3.10$) messages on average.
To compare the quality of feedback messages before and after the edits, we take both \textit{incorrect feedback} and \textit{shallow feedback} as \textit{low-quality feedback}. After that, there would be four types of quality change: low-to-low, low-to-high, high-to-low, and high-to-high. While the number of edits is similar for the two conditions, in the sampled edited dataset, there are $22.22\%$ low-to-high and $77.78\%$ high-to-high changes under \sys{} condition, while there are $45.00\%$ low-to-low, $20.00\%$ low-to-high and $5.00\%$ high-to-low, and $30.00\%$ high-to-high changes under the baseline condition. Moreover, using the Fisher Exact Probability Test~\cite{raymond1995exact}, we found that participants were significantly more likely to convert lower-quality feedback (i.e., \textit{incorrect feedback} and \textit{shallow feedback}) into higher-quality feedback ($p=0.001$).

\subsubsection{Perceived load in real-time system}
Participants filled the NASA-TLX~\cite{hart1988development} question (Table~\ref{tbl:nasa-tlx}) and rated the ease of use (\sys{}:$Median: 5$; baseline: $Median: 6$) and usefulness (\sys{}:$Median: 6$; baseline: $Median: 6.5$) of the systems. We did not obtain significant differences in the task loads, ease of use, and usefulness, which is not surprising, given that handling issues in a real-time large class is a naturally heavy load task for humanity. For instance, P12 suggested \textit{``the most mentally demand task is to manage the incoming changes of student's code activity and issues.''}. The high task loads also indicate an authentic simulation was achieved in both conditions, while the ease of use and usefulness level speak that the systems under both conditions are powerful enough, which meets our expectation of powering the system with the state-of-the-art LLMs.

Regarding the perceived ease of reviewing feedback, we could not observe a significant difference in ratings between the two conditions (\sys{}: $Median = 5.50, \sigma=1.93$; Baseline: $Median = 5.00, \sigma=1.55$). Most of the participants using \sys{} (P1, P3, P5, P7, P9, P11, P13, P15, P17) mentioned that the visual augmentation makes the review process\textit{ ``easy''} by \textit{``helping me like, break down the sections and go to the part that I needed to see easier''} (P1). Participants also found the color coding ``useful'' since \textit{`` I can directly check what the problem is and connect it to the feedback. For example, some of my rewriting processes are based on the highlighted code. If a line is highlighted as incorrect, I can optimize the AI-generated feedback directly.''}(P5). Participants also reported that the visual augmentation made things like \textit{``basic outline of what the feedback is going to say'' ``easy to understand''} and \textit{``visually clear''} (P11).

\subsubsection{Critical Issue Recommendation}
Participants using both conditions expressed high-level confidence in identifying struggling students (\sys{}: $Median = 6.00, \sigma=0.87$; Baseline: $Median = 6.00, \sigma=1.29$). Participants mentioned that \textit{``I did not find even a single one where it was something that I needed to edit a lot when it was talking about something completely unrelated to the problem.''(P19)}. Participants also states that the description of the critical issues \textit{`` decreases your cognitive load, (and) you start knowing the problem and have an idea of the feedback you want to give.''(P13)}. This shows that participants have high confidence in the critical issue classification which helps them give a general idea of feedback to give these students.

\subsubsection{Visualizing feedback for all students can still be overwhelming at times}
Despite having two-level clustering for instructors to inspect feedback for student's with similar issues, 4 out of 10 participants still felt overwhelmed while inspecting feedback using \sys{}. In particular, 2 participants mentioned the length of the feedback as the main challenge in quickly validating the the accuracy of the feedback
\textit{``If I have to really review the feedback so much, I could, you know, equivalently provide the same quality and quantity of feedback in a shorter time that meets my standard.(P8)''}.
Despite the effort of mitigating this shortcoming through color highlighting of code structure, which is well received by all participants using \sys{} in \textit{``check what the problem is and connect it to the feedback.'' (P5)}, instructors still often resort to reading feedback in detail before sending it to students.

\section{\MakeUppercase{discussion}}
In this section, we attempt to explain the reasons that led to our findings and discuss their connection to the existing literature.

\subsection{\sys{}'s participants are more strategic in steering LLM feedback generation} Participants in both conditions found LLM feedback generation easy to control, with similar frequencies of modifying feedback prompts across tasks . However, the nature of these modifications differed significantly. Baseline participants primarily adjusted feedback length, while \sys{} participants made more substantial alterations to overall generation strategy (e.g., omitting strategy generation). This behavioral difference led to measurable benefits: \sys{} participants had fewer changes made in the downstream feedback review task, while not devoting additional time to the feedback generation task.
Qualitatively, 8 out 10 \sys{} participants expressed positive opinions about the approach, compared to 2 out of 10 baseline participants. These findings suggest that \sys{}'s component-based feedback generation design encourages a more strategic feedback creation process. While increased control over LLM responses aligns with prior research on the impact of effective prompt engineering tools ~\cite{jiang2022promptmaker,Mishra2023PromptAidPE,Zamfirescu2023Why}, our study provides an additional insight: even when participants perceive similar levels of control, a component-based design can trigger more strategy-level changes in behavior and outcomes. This finding extends our understanding of LLM-assisted feedback systems, demonstrating that thoughtful interface design can not only facilitate control but also promote more effective use of AI-generated content in educational contexts.

\subsection{\sys{}'s feedback enhances instructors' understanding of students' mistakes and improves personalized feedback}

Instructors often encounter ``expert blind spots'' when assessing students' errors ~\cite{Nathan2003ExpertBS,Nathan2001ExpertBS}. This occurs when instructors, due to their advanced knowledge, struggle to fully recognize the causes and types of difficulties that novice learners face ~\cite{Borko1992LearningTT}. Our study revealed that \sys{}'s feedback helps instructors identify and correct these blind spots in two ways: by improving understanding of students' mistakes and by modeling more effective personalized feedback. For instance, in an exercise where students could use either a \texttt{Counter} or the length of a \texttt{List} to count elements, \sys{} suggested using \texttt{.append()} for the \texttt{List} approach. One participant (P15) initially thought this incorrect, not recognizing the alternative solution, but later adopted it after reflection. This demonstrates how \sys{} exposes instructors to things that they might not recognize.
Furthermore, \sys{}'s feedback is typically characterized by small, targeted hints (e.g., suggesting \texttt{.append()}) rather than those that include more than what the students needed. We saw that instructors acknowledged such concise and personalization aspects of the generated feedback and expressed how they should refine their own feedback strategies in the future, moving from overly comprehensive hints to more focused, relevant guidance. By shaping instructors' mental models and feedback strategies, our finding indicates that \sys{}'s way of generating feedback can also serve as a learning tool for instructors themselves to catch their blind spots.

\subsection{Visual binding improves the efficiency and accuracy of feedback repair}

Repairing lower-quality LLM-generated feedback is also a crucial part of the review process, but it can be challenging and inefficient. Our study revealed that \sys{}'s visual binding feature significantly enhances both the efficiency and accuracy of these repairs.
In the baseline condition, even when participants identified lower-quality feedback, their edits often failed to fully address all issues, resulting in partially improved but still suboptimal feedback. In contrast, \sys{} participants achieved a 100\% successful conversion rate in repairing low-quality feedback. 
Further investigation showed that baseline participants often overlooked issues in unedited portions of the feedback, such as the feedback is addressing an issue that the student does not have.
The visual binding feature in \sys{} enabled participants to notice these issues more accurately and comprehensively, preventing such oversights. 

\section{\MakeUppercase{limitations and future work}}

While our evaluation was done in a lab setting instead of in a real classroom -- and thus, a reduced set of distractions (i.e., the cognitive demands of being in a classroom setting) may impact the available attention that the instructor has to validate and generate feedback -- we believe it clearly shows that the overall net effect of \sys{} is positive. 
Our focus was on the impact of \sys{} on instructors' ability to provide feedback that they felt was effective, but we did not explore the impact on students directly. Future work will explore how different versions of feedback affect both perceived and real student outcomes in the medium and long term, and how this feedback can be fed back to instructors to give them deeper insights and enable more comprehensive personalization.

Future research will see \sys{} deployed in live classroom settings, which will provide a better understanding of how the use of AI-driven feedback tools works for all parties in an end-to-end application. The overall speed at which feedback can be provided in real settings will also impact how students use it. For example, one participant (P1) mentioned that it is possible that by the time they have finished crafting the feedback, the student might not have the same issue anymore. While our results demonstrate that instructors can provide better feedback in the same amount of time, per-student latency will still be a limiting factor, since instructor effort is serialized. 

Further, the time it takes current LLM models to generate feedback can be relatively long, especially in large classroom settings with many students and interactions to consider. While this may be feasible for larger institutions with significant compute resources (or those supported by donated educational resources from large technology companies), it is a limitation on current-day deployment of such tools. We expect that future advances will dramatically reduce the compute cost of run-time inference for LLMs, as we have seen with nearly all such computational bottlenecks in the past.
\section{\MakeUppercase{conclusion}}

In this paper, we present \sys{}, an interactive system designed to help programming instructors create personalized feedback at scale for class exercises that include writing code and discussing code in groups. 
By integrating LLM-powered issue detection and feedback generation with a structured review process for instructors, \sys{} significantly enhances instructors' ability to provide high-quality, personalized feedback at scale.
In a between-subject study, we showed that participants could use \sys{} to create more high-quality feedback compared to a baseline system with a standard linear conversational structure, as well as helping them be able to more frequently transform initially low-quality feedback into higher-quality versions.
Importantly, these improvements were achieved without a significant increase in required time, indicating a low net interaction cost. 
Beyond this, we found that \sys{} could improve instructors' understanding of students' mistakes, help identify and correct expert blind spots, and maintain higher engagement throughout the feedback process. 
These findings contribute new insights into the design of AI-assisted educational tools and open new avenues for scaling personalized feedback in programming education, addressing the challenges of real-time response, issue prioritization, and large-scale personalization while deepening instructors' engagement with feedback creation.

% \begin{acks}
% \end{acks}

\bibliographystyle{ACM-Reference-Format}
\bibliography{ref}


\begin{thebibliography}{52}

%%% the Web address.

\ifx \showCODEN    \undefined \def \showCODEN     #1{\unskip}     \fi
\ifx \showDOI      \undefined \def \showDOI       #1{#1}\fi
\ifx \showISBNx    \undefined \def \showISBNx     #1{\unskip}     \fi
\ifx \showISBNxiii \undefined \def \showISBNxiii  #1{\unskip}     \fi
\ifx \showISSN     \undefined \def \showISSN      #1{\unskip}     \fi
\ifx \showLCCN     \undefined \def \showLCCN      #1{\unskip}     \fi
\ifx \shownote     \undefined \def \shownote      #1{#1}          \fi
\ifx \showarticletitle \undefined \def \showarticletitle #1{#1}   \fi
\ifx \showURL      \undefined \def \showURL       {\relax}        \fi
% invisible to TeX
\providecommand\bibfield[2]{#2}
\providecommand\bibinfo[2]{#2}
\providecommand\natexlab[1]{#1}
\providecommand\showeprint[2][]{arXiv:#2}

\bibitem[Ahuja et~al\mbox{.}(2019)]%
        {ahuja2019edusense}
\bibfield{author}{\bibinfo{person}{Karan Ahuja}, \bibinfo{person}{Dohyun Kim}, \bibinfo{person}{Franceska Xhakaj}, \bibinfo{person}{Virag Varga}, \bibinfo{person}{Anne Xie}, \bibinfo{person}{Stanley Zhang}, \bibinfo{person}{Jay~Eric Townsend}, \bibinfo{person}{Chris Harrison}, \bibinfo{person}{Amy Ogan}, {and} \bibinfo{person}{Yuvraj Agarwal}.} \bibinfo{year}{2019}\natexlab{}.
\newblock \showarticletitle{EduSense: Practical classroom sensing at Scale}.
\newblock \bibinfo{journal}{\emph{Proceedings of the ACM on Interactive, Mobile, Wearable and Ubiquitous Technologies}} \bibinfo{volume}{3}, \bibinfo{number}{3} (\bibinfo{year}{2019}), \bibinfo{pages}{1--26}.
\newblock


\bibitem[Aleven et~al\mbox{.}(2022)]%
        {aleven2022dashboard}
\bibfield{author}{\bibinfo{person}{Vincent Aleven}, \bibinfo{person}{Jori Blankestijn}, \bibinfo{person}{LuEttaMae Lawrence}, \bibinfo{person}{Tomohiro Nagashima}, {and} \bibinfo{person}{Niels Taatgen}.} \bibinfo{year}{2022}\natexlab{}.
\newblock \showarticletitle{A dashboard to support teachers during students’ self-paced AI-supported problem-solving practice}. In \bibinfo{booktitle}{\emph{European Conference on Technology Enhanced Learning}}. Springer, \bibinfo{pages}{16--30}.
\newblock


\bibitem[Bender et~al\mbox{.}(2021)]%
        {Bender2021OnTD}
\bibfield{author}{\bibinfo{person}{Emily~M. Bender}, \bibinfo{person}{Timnit Gebru}, \bibinfo{person}{Angelina McMillan-Major}, {and} \bibinfo{person}{Shmargaret Shmitchell}.} \bibinfo{year}{2021}\natexlab{}.
\newblock \showarticletitle{On the Dangers of Stochastic Parrots: Can Language Models Be Too Big?}
\newblock \bibinfo{journal}{\emph{Proceedings of the 2021 ACM Conference on Fairness, Accountability, and Transparency}} (\bibinfo{year}{2021}).
\newblock
\urldef\tempurl%
\url{https://api.semanticscholar.org/CorpusID:262580630}
\showURL{%
\tempurl}


\bibitem[Borko et~al\mbox{.}(1992)]%
        {Borko1992LearningTT}
\bibfield{author}{\bibinfo{person}{Hilda Borko}, \bibinfo{person}{Margaret Eisenhart}, \bibinfo{person}{Catherine~A. Brown}, \bibinfo{person}{Robert~G. Underhill}, \bibinfo{person}{Doug Jones}, {and} \bibinfo{person}{Patricia~C. Agard}.} \bibinfo{year}{1992}\natexlab{}.
\newblock \showarticletitle{Learning to Teach Hard Mathematics: Do Novice Teachers and Their Instructors Give Up Too Easily?}
\newblock \bibinfo{journal}{\emph{Journal for Research in Mathematics Education}}  \bibinfo{volume}{23} (\bibinfo{year}{1992}), \bibinfo{pages}{194--222}.
\newblock
\urldef\tempurl%
\url{https://api.semanticscholar.org/CorpusID:227291162}
\showURL{%
\tempurl}


\bibitem[Brown et~al\mbox{.}(2020)]%
        {NEURIPS2020_1457c0d6}
\bibfield{author}{\bibinfo{person}{Tom Brown}, \bibinfo{person}{Benjamin Mann}, \bibinfo{person}{Nick Ryder}, \bibinfo{person}{Melanie Subbiah}, \bibinfo{person}{Jared~D Kaplan}, \bibinfo{person}{Prafulla Dhariwal}, \bibinfo{person}{Arvind Neelakantan}, \bibinfo{person}{Pranav Shyam}, \bibinfo{person}{Girish Sastry}, \bibinfo{person}{Amanda Askell}, \bibinfo{person}{Sandhini Agarwal}, \bibinfo{person}{Ariel Herbert-Voss}, \bibinfo{person}{Gretchen Krueger}, \bibinfo{person}{Tom Henighan}, \bibinfo{person}{Rewon Child}, \bibinfo{person}{Aditya Ramesh}, \bibinfo{person}{Daniel Ziegler}, \bibinfo{person}{Jeffrey Wu}, \bibinfo{person}{Clemens Winter}, \bibinfo{person}{Chris Hesse}, \bibinfo{person}{Mark Chen}, \bibinfo{person}{Eric Sigler}, \bibinfo{person}{Mateusz Litwin}, \bibinfo{person}{Scott Gray}, \bibinfo{person}{Benjamin Chess}, \bibinfo{person}{Jack Clark}, \bibinfo{person}{Christopher Berner}, \bibinfo{person}{Sam McCandlish}, \bibinfo{person}{Alec Radford}, \bibinfo{person}{Ilya Sutskever}, {and}
  \bibinfo{person}{Dario Amodei}.} \bibinfo{year}{2020}\natexlab{}.
\newblock \showarticletitle{Language Models are Few-Shot Learners}. In \bibinfo{booktitle}{\emph{Advances in Neural Information Processing Systems}}, \bibfield{editor}{\bibinfo{person}{H.~Larochelle}, \bibinfo{person}{M.~Ranzato}, \bibinfo{person}{R.~Hadsell}, \bibinfo{person}{M.F. Balcan}, {and} \bibinfo{person}{H.~Lin}} (Eds.), Vol.~\bibinfo{volume}{33}. \bibinfo{publisher}{Curran Associates, Inc.}, \bibinfo{pages}{1877--1901}.
\newblock
\urldef\tempurl%
\url{https://proceedings.neurips.cc/paper_files/paper/2020/file/1457c0d6bfcb4967418bfb8ac142f64a-Paper.pdf}
\showURL{%
\tempurl}


\bibitem[Butler and Winne(1995)]%
        {butler1995feedback}
\bibfield{author}{\bibinfo{person}{Deborah~L Butler} {and} \bibinfo{person}{Philip~H Winne}.} \bibinfo{year}{1995}\natexlab{}.
\newblock \showarticletitle{Feedback and self-regulated learning: A theoretical synthesis}.
\newblock \bibinfo{journal}{\emph{Review of educational research}} \bibinfo{volume}{65}, \bibinfo{number}{3} (\bibinfo{year}{1995}), \bibinfo{pages}{245--281}.
\newblock


\bibitem[{\c{C}}ardak and Selvi(2016)]%
        {ccardak2016increasing}
\bibfield{author}{\bibinfo{person}{{\c{C}}i{\u{g}}dem~Suzan {\c{C}}ardak} {and} \bibinfo{person}{K{\i}ymet Selvi}.} \bibinfo{year}{2016}\natexlab{}.
\newblock \showarticletitle{Increasing teacher candidates' ways of interaction and levels of learning through action research in a blended course}.
\newblock \bibinfo{journal}{\emph{Computers in Human Behavior}}  \bibinfo{volume}{61} (\bibinfo{year}{2016}), \bibinfo{pages}{488--506}.
\newblock


\bibitem[Chen et~al\mbox{.}(2020)]%
        {chen2020edcode}
\bibfield{author}{\bibinfo{person}{Yan Chen}, \bibinfo{person}{Jaylin Herskovitz}, \bibinfo{person}{Gabriel Matute}, \bibinfo{person}{April Wang}, \bibinfo{person}{Sang~Won Lee}, \bibinfo{person}{Walter~S Lasecki}, {and} \bibinfo{person}{Steve Oney}.} \bibinfo{year}{2020}\natexlab{}.
\newblock \showarticletitle{EdCode: Towards Personalized Support at Scale for Remote Assistance in CS Education}. In \bibinfo{booktitle}{\emph{2020 IEEE Symposium on Visual Languages and Human-Centric Computing (VL/HCC)}}. IEEE, \bibinfo{pages}{1--5}.
\newblock


\bibitem[Clow(2013)]%
        {clow2013overview}
\bibfield{author}{\bibinfo{person}{Doug Clow}.} \bibinfo{year}{2013}\natexlab{}.
\newblock \showarticletitle{An overview of learning analytics}.
\newblock \bibinfo{journal}{\emph{Teaching in Higher Education}} \bibinfo{volume}{18}, \bibinfo{number}{6} (\bibinfo{year}{2013}), \bibinfo{pages}{683--695}.
\newblock


\bibitem[Comas-Quinn(2011)]%
        {comas2011learning}
\bibfield{author}{\bibinfo{person}{Anna Comas-Quinn}.} \bibinfo{year}{2011}\natexlab{}.
\newblock \showarticletitle{Learning to teach online or learning to become an online teacher: An exploration of teachers’ experiences in a blended learning course}.
\newblock \bibinfo{journal}{\emph{ReCALL}} \bibinfo{volume}{23}, \bibinfo{number}{3} (\bibinfo{year}{2011}), \bibinfo{pages}{218--232}.
\newblock


\bibitem[Est{\'e}vez-Ayres et~al\mbox{.}(2024)]%
        {estevez2024evaluation}
\bibfield{author}{\bibinfo{person}{Iria Est{\'e}vez-Ayres}, \bibinfo{person}{Patricia Callejo}, \bibinfo{person}{Miguel~{\'A}ngel Hombrados-Herrera}, \bibinfo{person}{Carlos Alario-Hoyos}, {and} \bibinfo{person}{Carlos Delgado~Kloos}.} \bibinfo{year}{2024}\natexlab{}.
\newblock \showarticletitle{Evaluation of LLM Tools for Feedback Generation in a Course on Concurrent Programming}.
\newblock \bibinfo{journal}{\emph{International Journal of Artificial Intelligence in Education}} (\bibinfo{year}{2024}), \bibinfo{pages}{1--17}.
\newblock


\bibitem[Gabbay and Cohen(2024)]%
        {gabbay2024combining}
\bibfield{author}{\bibinfo{person}{Hagit Gabbay} {and} \bibinfo{person}{Anat Cohen}.} \bibinfo{year}{2024}\natexlab{}.
\newblock \showarticletitle{Combining LLM-Generated and Test-Based Feedback in a MOOC for Programming}. In \bibinfo{booktitle}{\emph{Proceedings of the Eleventh ACM Conference on Learning@ Scale}}. \bibinfo{pages}{177--187}.
\newblock


\bibitem[Gero et~al\mbox{.}(2024)]%
        {Gero2024Sensemaking}
\bibfield{author}{\bibinfo{person}{Katy~Ilonka Gero}, \bibinfo{person}{Chelse Swoopes}, \bibinfo{person}{Ziwei Gu}, \bibinfo{person}{Jonathan~K. Kummerfeld}, {and} \bibinfo{person}{Elena~L. Glassman}.} \bibinfo{year}{2024}\natexlab{}.
\newblock \showarticletitle{Supporting Sensemaking of Large Language Model Outputs at Scale}. In \bibinfo{booktitle}{\emph{Proceedings of the CHI Conference on Human Factors in Computing Systems}} (Honolulu, HI, USA) \emph{(\bibinfo{series}{CHI '24})}. \bibinfo{publisher}{Association for Computing Machinery}, \bibinfo{address}{New York, NY, USA}, Article \bibinfo{articleno}{838}, \bibinfo{numpages}{21}~pages.
\newblock
\showISBNx{9798400703300}
\urldef\tempurl%
\url{https://doi.org/10.1145/3613904.3642139}
\showDOI{\tempurl}


\bibitem[Glassman et~al\mbox{.}(2024)]%
        {glassman2024airesilientinterfaces}
\bibfield{author}{\bibinfo{person}{Elena~L. Glassman}, \bibinfo{person}{Ziwei Gu}, {and} \bibinfo{person}{Jonathan~K. Kummerfeld}.} \bibinfo{year}{2024}\natexlab{}.
\newblock \bibinfo{title}{AI-Resilient Interfaces}.
\newblock
\newblock
\showeprint[arxiv]{2405.08447}~[cs.HC]
\urldef\tempurl%
\url{https://arxiv.org/abs/2405.08447}
\showURL{%
\tempurl}


\bibitem[Greller and Drachsler(2012)]%
        {greller2012translating}
\bibfield{author}{\bibinfo{person}{Wolfgang Greller} {and} \bibinfo{person}{Hendrik Drachsler}.} \bibinfo{year}{2012}\natexlab{}.
\newblock \showarticletitle{Translating learning into numbers: A generic framework for learning analytics}.
\newblock \bibinfo{journal}{\emph{Journal of Educational Technology \& Society}} \bibinfo{volume}{15}, \bibinfo{number}{3} (\bibinfo{year}{2012}), \bibinfo{pages}{42--57}.
\newblock


\bibitem[Guo(2015)]%
        {guo2015codeopticon}
\bibfield{author}{\bibinfo{person}{Philip~J Guo}.} \bibinfo{year}{2015}\natexlab{}.
\newblock \showarticletitle{Codeopticon: Real-time, one-to-many human tutoring for computer programming}. In \bibinfo{booktitle}{\emph{Proceedings of the 28th Annual ACM Symposium on User Interface Software \& Technology}} (Charlotte, NC, USA) \emph{(\bibinfo{series}{UIST ’15})}. ACM, \bibinfo{publisher}{Association for Computing Machinery}, \bibinfo{address}{New York, NY, USA}, \bibinfo{pages}{599--608}.
\newblock
\showISBNx{9781450337793}
\urldef\tempurl%
\url{https://doi.org/10.1145/2807442.2807469}
\showDOI{\tempurl}


\bibitem[Hart and Staveland(1988)]%
        {hart1988development}
\bibfield{author}{\bibinfo{person}{Sandra~G Hart} {and} \bibinfo{person}{Lowell~E Staveland}.} \bibinfo{year}{1988}\natexlab{}.
\newblock \showarticletitle{Development of NASA-TLX (Task Load Index): Results of empirical and theoretical research}.
\newblock In \bibinfo{booktitle}{\emph{Advances in psychology}}. Vol.~\bibinfo{volume}{52}. \bibinfo{publisher}{Elsevier}, \bibinfo{pages}{139--183}.
\newblock


\bibitem[Hattie(2012)]%
        {hattie2012visible}
\bibfield{author}{\bibinfo{person}{John Hattie}.} \bibinfo{year}{2012}\natexlab{}.
\newblock \bibinfo{booktitle}{\emph{Visible learning for teachers: Maximizing impact on learning}}.
\newblock \bibinfo{publisher}{Routledge}.
\newblock


\bibitem[Hattie and Timperley(2007)]%
        {hattie2007power}
\bibfield{author}{\bibinfo{person}{John Hattie} {and} \bibinfo{person}{Helen Timperley}.} \bibinfo{year}{2007}\natexlab{}.
\newblock \showarticletitle{The power of feedback}.
\newblock \bibinfo{journal}{\emph{Review of educational research}} \bibinfo{volume}{77}, \bibinfo{number}{1} (\bibinfo{year}{2007}), \bibinfo{pages}{81--112}.
\newblock


\bibitem[Holstein et~al\mbox{.}(2018)]%
        {holstein2018classroom}
\bibfield{author}{\bibinfo{person}{Kenneth Holstein}, \bibinfo{person}{Gena Hong}, \bibinfo{person}{Mera Tegene}, \bibinfo{person}{Bruce~M McLaren}, {and} \bibinfo{person}{Vincent Aleven}.} \bibinfo{year}{2018}\natexlab{}.
\newblock \showarticletitle{The classroom as a dashboard: Co-designing wearable cognitive augmentation for K-12 teachers}. In \bibinfo{booktitle}{\emph{Proceedings of the 8th international conference on learning Analytics and knowledge}}. \bibinfo{pages}{79--88}.
\newblock


\bibitem[Huang et~al\mbox{.}(2023)]%
        {huang2023surveyhallucinationlargelanguage}
\bibfield{author}{\bibinfo{person}{Lei Huang}, \bibinfo{person}{Weijiang Yu}, \bibinfo{person}{Weitao Ma}, \bibinfo{person}{Weihong Zhong}, \bibinfo{person}{Zhangyin Feng}, \bibinfo{person}{Haotian Wang}, \bibinfo{person}{Qianglong Chen}, \bibinfo{person}{Weihua Peng}, \bibinfo{person}{Xiaocheng Feng}, \bibinfo{person}{Bing Qin}, {and} \bibinfo{person}{Ting Liu}.} \bibinfo{year}{2023}\natexlab{}.
\newblock \bibinfo{title}{A Survey on Hallucination in Large Language Models: Principles, Taxonomy, Challenges, and Open Questions}.
\newblock
\newblock
\showeprint[arxiv]{2311.05232}~[cs.CL]
\urldef\tempurl%
\url{https://arxiv.org/abs/2311.05232}
\showURL{%
\tempurl}


\bibitem[Jiang et~al\mbox{.}(2022)]%
        {jiang2022promptmaker}
\bibfield{author}{\bibinfo{person}{Ellen Jiang}, \bibinfo{person}{Kristen Olson}, \bibinfo{person}{Edwin Toh}, \bibinfo{person}{Alejandra Molina}, \bibinfo{person}{Aaron Donsbach}, \bibinfo{person}{Michael Terry}, {and} \bibinfo{person}{Carrie~J Cai}.} \bibinfo{year}{2022}\natexlab{}.
\newblock \showarticletitle{PromptMaker: Prompt-based Prototyping with Large Language Models}. In \bibinfo{booktitle}{\emph{Extended Abstracts of the 2022 CHI Conference on Human Factors in Computing Systems}} (New Orleans, LA, USA) \emph{(\bibinfo{series}{CHI EA '22})}. \bibinfo{publisher}{Association for Computing Machinery}, \bibinfo{address}{New York, NY, USA}, Article \bibinfo{articleno}{35}, \bibinfo{numpages}{8}~pages.
\newblock
\showISBNx{9781450391566}
\urldef\tempurl%
\url{https://doi.org/10.1145/3491101.3503564}
\showDOI{\tempurl}


\bibitem[Jung and Wise(2024)]%
        {jung2024probing}
\bibfield{author}{\bibinfo{person}{Yeonji Jung} {and} \bibinfo{person}{Alyssa~Friend Wise}.} \bibinfo{year}{2024}\natexlab{}.
\newblock \showarticletitle{Probing Actionability in Learning Analytics: The Role of Routines, Timing, and Pathways}. In \bibinfo{booktitle}{\emph{Proceedings of the 14th Learning Analytics and Knowledge Conference}}. \bibinfo{pages}{871--877}.
\newblock


\bibitem[Karademir et~al\mbox{.}(2021)]%
        {karademir2024LACockpit}
\bibfield{author}{\bibinfo{person}{Onur Karademir}, \bibinfo{person}{Atezaz Ahmad}, \bibinfo{person}{Jan Schneider}, \bibinfo{person}{Daniele {Di Mitri}}, \bibinfo{person}{Ioana Jivet}, {and} \bibinfo{person}{Hendrik Drachsler}.} \bibinfo{year}{2021}\natexlab{}.
\newblock \showarticletitle{Designing the Learning Analytics Cockpit - A Dashboard that Enables Interventions}. In \bibinfo{booktitle}{\emph{Methodologies and Intelligent Systems for Technology Enhanced Learning, 11th International Conference}} \emph{(\bibinfo{series}{Lecture Notes in Networks and Systems})}, \bibfield{editor}{\bibinfo{person}{Fernando {De la Prieta}}, \bibinfo{person}{Rosella Gennari}, \bibinfo{person}{Marco Temperini}, \bibinfo{person}{Tania {Di Mascio}}, \bibinfo{person}{Pierpaolo Vittorini}, \bibinfo{person}{Zuzana Kubincova}, \bibinfo{person}{Elvira Popescu}, \bibinfo{person}{Davide {Rua Carneiro}}, \bibinfo{person}{Loreto Lancia}, {and} \bibinfo{person}{Agnese Addone}} (Eds.). \bibinfo{publisher}{Springer}, \bibinfo{pages}{95--104}.
\newblock
\showISBNx{9783030866174}
\urldef\tempurl%
\url{https://doi.org/10.1007/978-3-030-86618-1_10}
\showDOI{\tempurl}
\newblock
\shownote{11th International Conference on Methodologies and Intelligent Systems for Technology Enhanced Learning, MIS4TEL 2021 ; Conference date: 06-10-2021 Through 08-10-2021}.


\bibitem[Karademir et~al\mbox{.}(2024)]%
        {karademir2024don}
\bibfield{author}{\bibinfo{person}{Onur Karademir}, \bibinfo{person}{Daniele Di~Mitri}, \bibinfo{person}{Jan Schneider}, \bibinfo{person}{Ioana Jivet}, \bibinfo{person}{J{\"o}rn Allmang}, \bibinfo{person}{Sebastian Gombert}, \bibinfo{person}{Marcus Kubsch}, \bibinfo{person}{Knut Neumann}, {and} \bibinfo{person}{Hendrik Drachsler}.} \bibinfo{year}{2024}\natexlab{}.
\newblock \showarticletitle{I don't have time! But keep me in the loop: Co-designing requirements for a learning analytics cockpit with teachers}.
\newblock \bibinfo{journal}{\emph{Journal of Computer Assisted Learning}} (\bibinfo{year}{2024}).
\newblock


\bibitem[Ma et~al\mbox{.}(2022)]%
        {ma2022glancee}
\bibfield{author}{\bibinfo{person}{Shuai Ma}, \bibinfo{person}{Taichang Zhou}, \bibinfo{person}{Fei Nie}, {and} \bibinfo{person}{Xiaojuan Ma}.} \bibinfo{year}{2022}\natexlab{}.
\newblock \showarticletitle{Glancee: An adaptable system for instructors to grasp student learning status in synchronous online classes}. In \bibinfo{booktitle}{\emph{Proceedings of the 2022 CHI Conference on Human Factors in Computing Systems}}. \bibinfo{pages}{1--25}.
\newblock


\bibitem[McDonald et~al\mbox{.}(2019)]%
        {10.1145/3359174}
\bibfield{author}{\bibinfo{person}{Nora McDonald}, \bibinfo{person}{Sarita Schoenebeck}, {and} \bibinfo{person}{Andrea Forte}.} \bibinfo{year}{2019}\natexlab{}.
\newblock \showarticletitle{Reliability and Inter-rater Reliability in Qualitative Research: Norms and Guidelines for CSCW and HCI Practice}.
\newblock \bibinfo{journal}{\emph{Proc. ACM Hum.-Comput. Interact.}} \bibinfo{volume}{3}, \bibinfo{number}{CSCW}, Article \bibinfo{articleno}{72} (\bibinfo{date}{nov} \bibinfo{year}{2019}), \bibinfo{numpages}{23}~pages.
\newblock
\urldef\tempurl%
\url{https://doi.org/10.1145/3359174}
\showDOI{\tempurl}


\bibitem[Mishra et~al\mbox{.}(2023)]%
        {Mishra2023PromptAidPE}
\bibfield{author}{\bibinfo{person}{Aditi Mishra}, \bibinfo{person}{Utkarsh Soni}, \bibinfo{person}{Anjana Arunkumar}, \bibinfo{person}{Jinbin Huang}, \bibinfo{person}{Bum~Chul Kwon}, {and} \bibinfo{person}{Chris Bryan}.} \bibinfo{year}{2023}\natexlab{}.
\newblock \showarticletitle{PromptAid: Prompt Exploration, Perturbation, Testing and Iteration using Visual Analytics for Large Language Models}.
\newblock \bibinfo{journal}{\emph{ArXiv}}  \bibinfo{volume}{abs/2304.01964} (\bibinfo{year}{2023}).
\newblock
\urldef\tempurl%
\url{https://api.semanticscholar.org/CorpusID:257921397}
\showURL{%
\tempurl}


\bibitem[Murali et~al\mbox{.}(2021)]%
        {murali2021affectivespotlight}
\bibfield{author}{\bibinfo{person}{Prasanth Murali}, \bibinfo{person}{Javier Hernandez}, \bibinfo{person}{Daniel McDuff}, \bibinfo{person}{Kael Rowan}, \bibinfo{person}{Jina Suh}, {and} \bibinfo{person}{Mary Czerwinski}.} \bibinfo{year}{2021}\natexlab{}.
\newblock \showarticletitle{Affectivespotlight: Facilitating the communication of affective responses from audience members during online presentations}. In \bibinfo{booktitle}{\emph{Proceedings of the 2021 CHI Conference on Human Factors in Computing Systems}}. \bibinfo{pages}{1--13}.
\newblock


\bibitem[Nam et~al\mbox{.}(2024)]%
        {nam2024using}
\bibfield{author}{\bibinfo{person}{Daye Nam}, \bibinfo{person}{Andrew Macvean}, \bibinfo{person}{Vincent Hellendoorn}, \bibinfo{person}{Bogdan Vasilescu}, {and} \bibinfo{person}{Brad Myers}.} \bibinfo{year}{2024}\natexlab{}.
\newblock \showarticletitle{Using an llm to help with code understanding}. In \bibinfo{booktitle}{\emph{Proceedings of the IEEE/ACM 46th International Conference on Software Engineering}}. \bibinfo{pages}{1--13}.
\newblock


\bibitem[Nathan et~al\mbox{.}(2001)]%
        {Nathan2001ExpertBS}
\bibfield{author}{\bibinfo{person}{Mitchell~J. Nathan}, \bibinfo{person}{K. Koedinger}, {and} \bibinfo{person}{Martha~W. Alibali}.} \bibinfo{year}{2001}\natexlab{}.
\newblock \showarticletitle{Expert Blind Spot : When Content Knowledge Eclipses Pedagogical Content Knowledge}.
\newblock
\urldef\tempurl%
\url{https://api.semanticscholar.org/CorpusID:14779203}
\showURL{%
\tempurl}


\bibitem[Nathan and Petrosino(2003)]%
        {Nathan2003ExpertBS}
\bibfield{author}{\bibinfo{person}{Mitchell~J. Nathan} {and} \bibinfo{person}{Anthony~J. Petrosino}.} \bibinfo{year}{2003}\natexlab{}.
\newblock \showarticletitle{Expert Blind Spot Among Preservice Teachers}.
\newblock \bibinfo{journal}{\emph{American Educational Research Journal}}  \bibinfo{volume}{40} (\bibinfo{year}{2003}), \bibinfo{pages}{905 -- 928}.
\newblock
\urldef\tempurl%
\url{https://api.semanticscholar.org/CorpusID:145129059}
\showURL{%
\tempurl}


\bibitem[Ngoon et~al\mbox{.}(2024)]%
        {ngoon2024classinsights}
\bibfield{author}{\bibinfo{person}{Tricia~J. Ngoon}, \bibinfo{person}{S Sushil}, \bibinfo{person}{Angela~E.B. Stewart}, \bibinfo{person}{Ung-Sang Lee}, \bibinfo{person}{Saranya Venkatraman}, \bibinfo{person}{Neil Thawani}, \bibinfo{person}{Prasenjit Mitra}, \bibinfo{person}{Sherice Clarke}, \bibinfo{person}{John Zimmerman}, {and} \bibinfo{person}{Amy Ogan}.} \bibinfo{year}{2024}\natexlab{}.
\newblock \showarticletitle{ClassInSight: Designing Conversation Support Tools to Visualize Classroom Discussion for Personalized Teacher Professional Development}. In \bibinfo{booktitle}{\emph{Proceedings of the CHI Conference on Human Factors in Computing Systems}} (Honolulu, HI, USA) \emph{(\bibinfo{series}{CHI '24})}. \bibinfo{publisher}{Association for Computing Machinery}, \bibinfo{address}{New York, NY, USA}, Article \bibinfo{articleno}{663}, \bibinfo{numpages}{15}~pages.
\newblock
\showISBNx{9798400703300}
\urldef\tempurl%
\url{https://doi.org/10.1145/3613904.3642487}
\showDOI{\tempurl}


\bibitem[Nguyen et~al\mbox{.}(2024)]%
        {nguyen2024beginning}
\bibfield{author}{\bibinfo{person}{Sydney Nguyen}, \bibinfo{person}{Hannah~McLean Babe}, \bibinfo{person}{Yangtian Zi}, \bibinfo{person}{Arjun Guha}, \bibinfo{person}{Carolyn~Jane Anderson}, {and} \bibinfo{person}{Molly~Q Feldman}.} \bibinfo{year}{2024}\natexlab{}.
\newblock \showarticletitle{How Beginning Programmers and Code LLMs (Mis) read Each Other}. In \bibinfo{booktitle}{\emph{Proceedings of the CHI Conference on Human Factors in Computing Systems}}. \bibinfo{pages}{1--26}.
\newblock


\bibitem[Raymond and Rousset(1995)]%
        {raymond1995exact}
\bibfield{author}{\bibinfo{person}{Michel Raymond} {and} \bibinfo{person}{Fran{\c{c}}ois Rousset}.} \bibinfo{year}{1995}\natexlab{}.
\newblock \showarticletitle{An exact test for population differentiation}.
\newblock \bibinfo{journal}{\emph{Evolution}} (\bibinfo{year}{1995}), \bibinfo{pages}{1280--1283}.
\newblock


\bibitem[Robins et~al\mbox{.}(2003)]%
        {Robins2003LearningAT}
\bibfield{author}{\bibinfo{person}{Anthony~V. Robins}, \bibinfo{person}{Janet Rountree}, {and} \bibinfo{person}{Nathan Rountree}.} \bibinfo{year}{2003}\natexlab{}.
\newblock \showarticletitle{Learning and Teaching Programming: A Review and Discussion}.
\newblock \bibinfo{journal}{\emph{Computer Science Education}}  \bibinfo{volume}{13} (\bibinfo{year}{2003}), \bibinfo{pages}{137 -- 172}.
\newblock
\urldef\tempurl%
\url{https://api.semanticscholar.org/CorpusID:10565822}
\showURL{%
\tempurl}


\bibitem[Sato et~al\mbox{.}(2023)]%
        {sato2023groupnamics}
\bibfield{author}{\bibinfo{person}{Arissa~J Sato}, \bibinfo{person}{Zefan Sramek}, {and} \bibinfo{person}{Koji Yatani}.} \bibinfo{year}{2023}\natexlab{}.
\newblock \showarticletitle{Groupnamics: Designing an interface for overviewing and managing parallel group discussions in an online classroom}. In \bibinfo{booktitle}{\emph{Proceedings of the 2023 CHI Conference on Human Factors in Computing Systems}}. \bibinfo{pages}{1--18}.
\newblock


\bibitem[Schneider et~al\mbox{.}(2021)]%
        {schneider2021collaboration}
\bibfield{author}{\bibinfo{person}{Bertrand Schneider}, \bibinfo{person}{Nia Dowell}, {and} \bibinfo{person}{Kate Thompson}.} \bibinfo{year}{2021}\natexlab{}.
\newblock \showarticletitle{Collaboration analytics—current state and potential futures}.
\newblock \bibinfo{journal}{\emph{Journal of Learning Analytics}} \bibinfo{volume}{8}, \bibinfo{number}{1} (\bibinfo{year}{2021}), \bibinfo{pages}{1--12}.
\newblock


\bibitem[Sridhar et~al\mbox{.}(2023)]%
        {Sridhar2023HarnessingLI}
\bibfield{author}{\bibinfo{person}{Pragnya Sridhar}, \bibinfo{person}{Aidan Doyle}, \bibinfo{person}{Arav Agarwal}, \bibinfo{person}{Chris Bogart}, \bibinfo{person}{Jaromir Savelka}, {and} \bibinfo{person}{Majd~F. Sakr}.} \bibinfo{year}{2023}\natexlab{}.
\newblock \showarticletitle{Harnessing LLMs in Curricular Design: Using GPT-4 to Support Authoring of Learning Objectives}. In \bibinfo{booktitle}{\emph{LLM@AIED}}.
\newblock
\urldef\tempurl%
\url{https://api.semanticscholar.org/CorpusID:259308858}
\showURL{%
\tempurl}


\bibitem[Suraworachet et~al\mbox{.}(2024)]%
        {10.1145/3636555.3636905}
\bibfield{author}{\bibinfo{person}{Wannapon Suraworachet}, \bibinfo{person}{Jennifer Seon}, {and} \bibinfo{person}{Mutlu Cukurova}.} \bibinfo{year}{2024}\natexlab{}.
\newblock \showarticletitle{Predicting challenge moments from students' discourse: A comparison of GPT-4 to two traditional natural language processing approaches}. In \bibinfo{booktitle}{\emph{Proceedings of the 14th Learning Analytics and Knowledge Conference}} (Kyoto, Japan) \emph{(\bibinfo{series}{LAK '24})}. \bibinfo{publisher}{Association for Computing Machinery}, \bibinfo{address}{New York, NY, USA}, \bibinfo{pages}{473–485}.
\newblock
\showISBNx{9798400716188}
\urldef\tempurl%
\url{https://doi.org/10.1145/3636555.3636905}
\showDOI{\tempurl}


\bibitem[Tang et~al\mbox{.}(2023)]%
        {tang2023vizpi}
\bibfield{author}{\bibinfo{person}{Xiaohang Tang}, \bibinfo{person}{Xi Chen}, \bibinfo{person}{Sam Wong}, {and} \bibinfo{person}{Yan Chen}.} \bibinfo{year}{2023}\natexlab{}.
\newblock \showarticletitle{VizPI: A Real-Time Visualization Tool for Enhancing Peer Instruction in Large-Scale Programming Lectures}. In \bibinfo{booktitle}{\emph{Adjunct Proceedings of the 36th Annual ACM Symposium on User Interface Software and Technology}} (San Francisco, CA, USA) \emph{(\bibinfo{series}{UIST '23 Adjunct})}. \bibinfo{publisher}{Association for Computing Machinery}, \bibinfo{address}{New York, NY, USA}, Article \bibinfo{articleno}{17}, \bibinfo{numpages}{3}~pages.
\newblock
\showISBNx{9798400700965}
\urldef\tempurl%
\url{https://doi.org/10.1145/3586182.3616632}
\showDOI{\tempurl}


\bibitem[Tang et~al\mbox{.}(2024)]%
        {tang2024vizgroup}
\bibfield{author}{\bibinfo{person}{Xiaohang Tang}, \bibinfo{person}{Sam Wong}, \bibinfo{person}{Kevin Pu}, \bibinfo{person}{Xi Chen}, \bibinfo{person}{Yalong Yang}, {and} \bibinfo{person}{Yan Chen}.} \bibinfo{year}{2024}\natexlab{}.
\newblock \showarticletitle{VizGroup: . VizGroup: An AI-Assisted Event-Driven System for Real-Time Collaborative Programming Learning Analytics.}, In \bibinfo{booktitle}{Proceedings of the 37th Annual ACM Symposium on User Interface Software \& Technology} (Pitts, PA, USA).
\newblock \bibinfo{journal}{\emph{arXiv preprint arXiv:2404.08743}}.
\newblock
\urldef\tempurl%
\url{https://doi.org/10.48550/arXiv.2404.08743}
\showDOI{\tempurl}
\showeprint[arxiv]{2404.08743}~[cs.HC]


\bibitem[Van~de Pol et~al\mbox{.}(2010)]%
        {van2010scaffolding}
\bibfield{author}{\bibinfo{person}{Janneke Van~de Pol}, \bibinfo{person}{Monique Volman}, {and} \bibinfo{person}{Jos Beishuizen}.} \bibinfo{year}{2010}\natexlab{}.
\newblock \showarticletitle{Scaffolding in teacher--student interaction: A decade of research}.
\newblock \bibinfo{journal}{\emph{Educational psychology review}}  \bibinfo{volume}{22} (\bibinfo{year}{2010}), \bibinfo{pages}{271--296}.
\newblock


\bibitem[van Leeuwen et~al\mbox{.}(2022)]%
        {van2022teacher}
\bibfield{author}{\bibinfo{person}{Anouschka van Leeuwen}, \bibinfo{person}{Stephanie~D Teasley}, {and} \bibinfo{person}{Alyssa~Friend Wise}.} \bibinfo{year}{2022}\natexlab{}.
\newblock \showarticletitle{Teacher and student facing learning analytics}.
\newblock \bibinfo{journal}{\emph{Handbook of learning analytics}} (\bibinfo{year}{2022}), \bibinfo{pages}{130--140}.
\newblock


\bibitem[Wang et~al\mbox{.}(2021)]%
        {wang2021puzzleme}
\bibfield{author}{\bibinfo{person}{April~Yi Wang}, \bibinfo{person}{Yan Chen}, \bibinfo{person}{John Joon~Young Chung}, \bibinfo{person}{Christopher Brooks}, {and} \bibinfo{person}{Steve Oney}.} \bibinfo{year}{2021}\natexlab{}.
\newblock \showarticletitle{PuzzleMe: Leveraging Peer Assessment for In-Class Programming Exercises}.
\newblock \bibinfo{journal}{\emph{Proceedings of the ACM on Human-Computer Interaction}} \bibinfo{volume}{5}, \bibinfo{number}{CSCW2} (\bibinfo{year}{2021}), \bibinfo{pages}{1--24}.
\newblock


\bibitem[Wise and Jung(2019)]%
        {wise2019teaching}
\bibfield{author}{\bibinfo{person}{Alyssa~Friend Wise} {and} \bibinfo{person}{Yeonji Jung}.} \bibinfo{year}{2019}\natexlab{}.
\newblock \showarticletitle{Teaching with analytics: Towards a situated model of instructional decision-making}.
\newblock \bibinfo{journal}{\emph{Journal of Learning Analytics}} \bibinfo{volume}{6}, \bibinfo{number}{2} (\bibinfo{year}{2019}), \bibinfo{pages}{53--69}.
\newblock


\bibitem[Wu et~al\mbox{.}(2024)]%
        {wu2024impact}
\bibfield{author}{\bibinfo{person}{Tong Wu}, \bibinfo{person}{Xiaohang Tang}, \bibinfo{person}{Sam Wong}, \bibinfo{person}{Xi Chen}, \bibinfo{person}{Clifford~A Shaffer}, {and} \bibinfo{person}{Yan Chen}.} \bibinfo{year}{2024}\natexlab{}.
\newblock \showarticletitle{The Impact of Group Discussion and Formation on Student Performance: An Experience Report in a Large CS1 Course}.
\newblock \bibinfo{journal}{\emph{arXiv preprint arXiv:2408.14610}} (\bibinfo{year}{2024}).
\newblock


\bibitem[Wu et~al\mbox{.}(2022)]%
        {10.1145/3491102.3517582}
\bibfield{author}{\bibinfo{person}{Tongshuang Wu}, \bibinfo{person}{Michael Terry}, {and} \bibinfo{person}{Carrie~Jun Cai}.} \bibinfo{year}{2022}\natexlab{}.
\newblock \showarticletitle{AI Chains: Transparent and Controllable Human-AI Interaction by Chaining Large Language Model Prompts}. In \bibinfo{booktitle}{\emph{Proceedings of the 2022 CHI Conference on Human Factors in Computing Systems}} (New Orleans, LA, USA) \emph{(\bibinfo{series}{CHI '22})}. \bibinfo{publisher}{Association for Computing Machinery}, \bibinfo{address}{New York, NY, USA}, Article \bibinfo{articleno}{385}, \bibinfo{numpages}{22}~pages.
\newblock
\showISBNx{9781450391573}
\urldef\tempurl%
\url{https://doi.org/10.1145/3491102.3517582}
\showDOI{\tempurl}


\bibitem[Yang et~al\mbox{.}(2023)]%
        {yang2023pair}
\bibfield{author}{\bibinfo{person}{Kexin~Bella Yang}, \bibinfo{person}{Vanessa Echeverria}, \bibinfo{person}{Zijing Lu}, \bibinfo{person}{Hongyu Mao}, \bibinfo{person}{Kenneth Holstein}, \bibinfo{person}{Nikol Rummel}, {and} \bibinfo{person}{Vincent Aleven}.} \bibinfo{year}{2023}\natexlab{}.
\newblock \showarticletitle{Pair-up: prototyping human-AI co-orchestration of dynamic transitions between individual and collaborative learning in the classroom}. In \bibinfo{booktitle}{\emph{Proceedings of the 2023 CHI conference on human factors in computing systems}}. \bibinfo{pages}{1--17}.
\newblock


\bibitem[Zamfirescu-Pereira et~al\mbox{.}(2023)]%
        {Zamfirescu2023Why}
\bibfield{author}{\bibinfo{person}{J.D. Zamfirescu-Pereira}, \bibinfo{person}{Richmond~Y. Wong}, \bibinfo{person}{Bjoern Hartmann}, {and} \bibinfo{person}{Qian Yang}.} \bibinfo{year}{2023}\natexlab{}.
\newblock \showarticletitle{Why Johnny Can’t Prompt: How Non-AI Experts Try (and Fail) to Design LLM Prompts}. In \bibinfo{booktitle}{\emph{Proceedings of the 2023 CHI Conference on Human Factors in Computing Systems}} (Hamburg, Germany) \emph{(\bibinfo{series}{CHI '23})}. \bibinfo{publisher}{Association for Computing Machinery}, \bibinfo{address}{New York, NY, USA}, Article \bibinfo{articleno}{437}, \bibinfo{numpages}{21}~pages.
\newblock
\showISBNx{9781450394215}
\urldef\tempurl%
\url{https://doi.org/10.1145/3544548.3581388}
\showDOI{\tempurl}


\bibitem[Zhang et~al\mbox{.}(2023)]%
        {zhang2023vizprog}
\bibfield{author}{\bibinfo{person}{Ashley~Ge Zhang}, \bibinfo{person}{Yan Chen}, {and} \bibinfo{person}{Steve Oney}.} \bibinfo{year}{2023}\natexlab{}.
\newblock \showarticletitle{VizProg: Identifying Misunderstandings By Visualizing Students’ Coding Progress}. In \bibinfo{booktitle}{\emph{Proceedings of the 2023 CHI Conference on Human Factors in Computing Systems}}. \bibinfo{pages}{1--16}.
\newblock


\bibitem[Zhang et~al\mbox{.}(2024)]%
        {zhang2024CFlow}
\bibfield{author}{\bibinfo{person}{Ashley~Ge Zhang}, \bibinfo{person}{Xiaohang Tang}, \bibinfo{person}{Steve Oney}, {and} \bibinfo{person}{Yan Chen}.} \bibinfo{year}{2024}\natexlab{}.
\newblock \showarticletitle{CFlow: Supporting Semantic Flow Analysis of Students' Code in Programming Problems at Scale}. In \bibinfo{booktitle}{\emph{Proceedings of the Eleventh ACM Conference on Learning @ Scale}} (Atlanta, GA, USA) \emph{(\bibinfo{series}{L@S '24})}. \bibinfo{publisher}{Association for Computing Machinery}, \bibinfo{address}{New York, NY, USA}, \bibinfo{pages}{188–199}.
\newblock
\showISBNx{9798400706332}
\urldef\tempurl%
\url{https://doi.org/10.1145/3657604.3662025}
\showDOI{\tempurl}


\end{thebibliography}

\renewcommand{\thesubsection}{\Alph{subsection}.}

\pagebreak
\section{\MakeUppercase{Appendix}}
\subsection{Appendix A: Annotator Interface}~\label{appendixA}
\label{sec:annotator A}
\begin{figure*} [!h]
    \centering
    \includegraphics[width=1\linewidth]{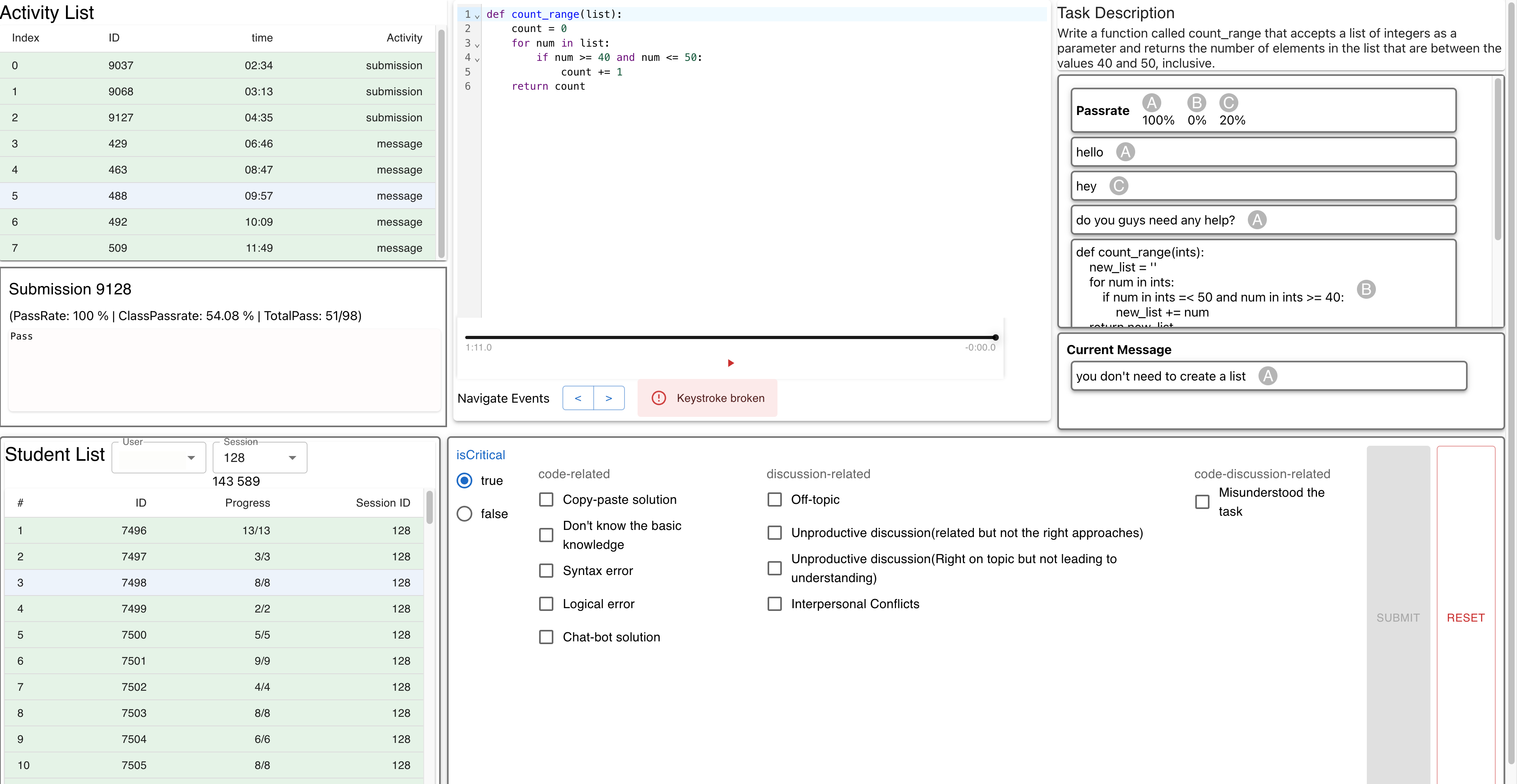}
    \caption{User Interface for annotator tool. Annotators select an activity from each student (Left) and provide a label by reviewing the playback for the selected activity.}
    \label{fig:annotator}
\end{figure*}

\pagebreak
\subsection{Appendix B: Feedback Annotator Interface}~\label{appendixB}
\begin{figure*} [!h]
    \centering
    \includegraphics[width=1\linewidth]{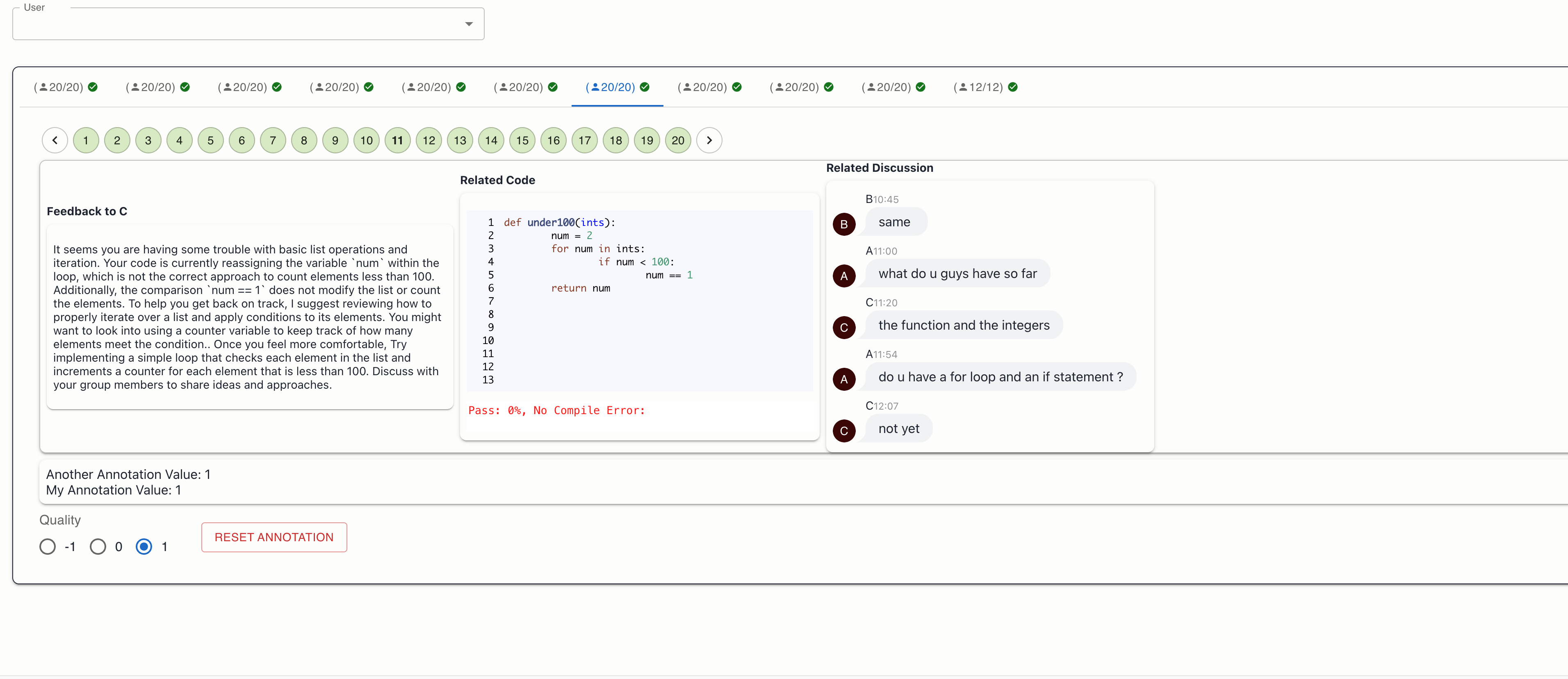}
    \caption{User Interface for annotating feedback. Annotators select the a feedback message from the message list (Top) and provide a label by reviewing the content of the message and related codes and discussions. }
    \label{fig:feedback_annotator}
\end{figure*}
\pagebreak
\subsection{Appendix C: Prompt for identifying  student's struggles}~\label{appendixC}
\begin{figure*} [!h]
    \centering
    \includegraphics[width=0.9\linewidth]{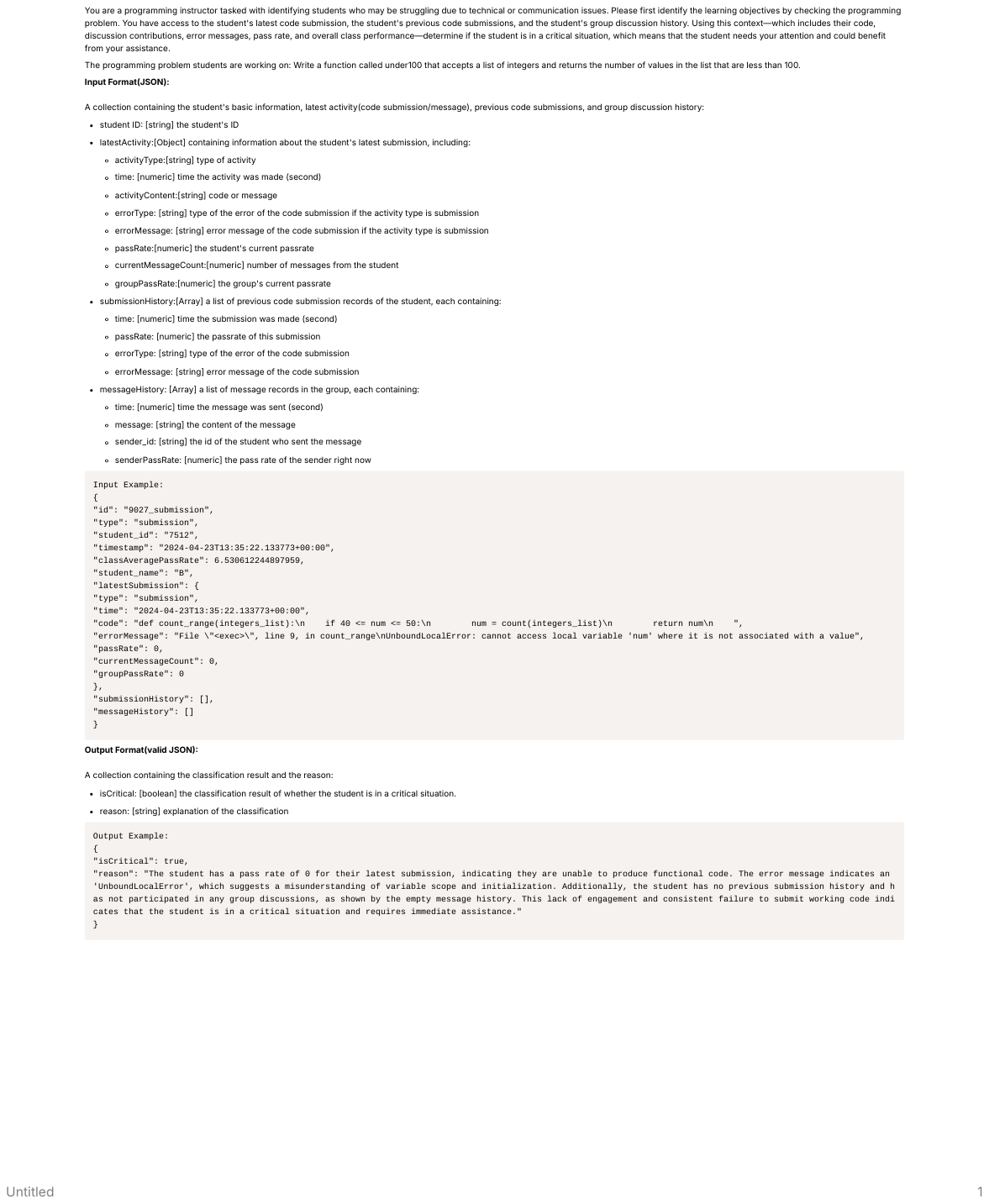}
    \caption{Prompt for identifying critical issue}
    \label{fig:critical_1}
\end{figure*}
\pagebreak
\subsection{Appendix D: Prompt for identifying critical issues in student's code}~\label{appendixD}
\begin{figure*} [!h]
    \centering
    \includegraphics[width=0.9\linewidth]{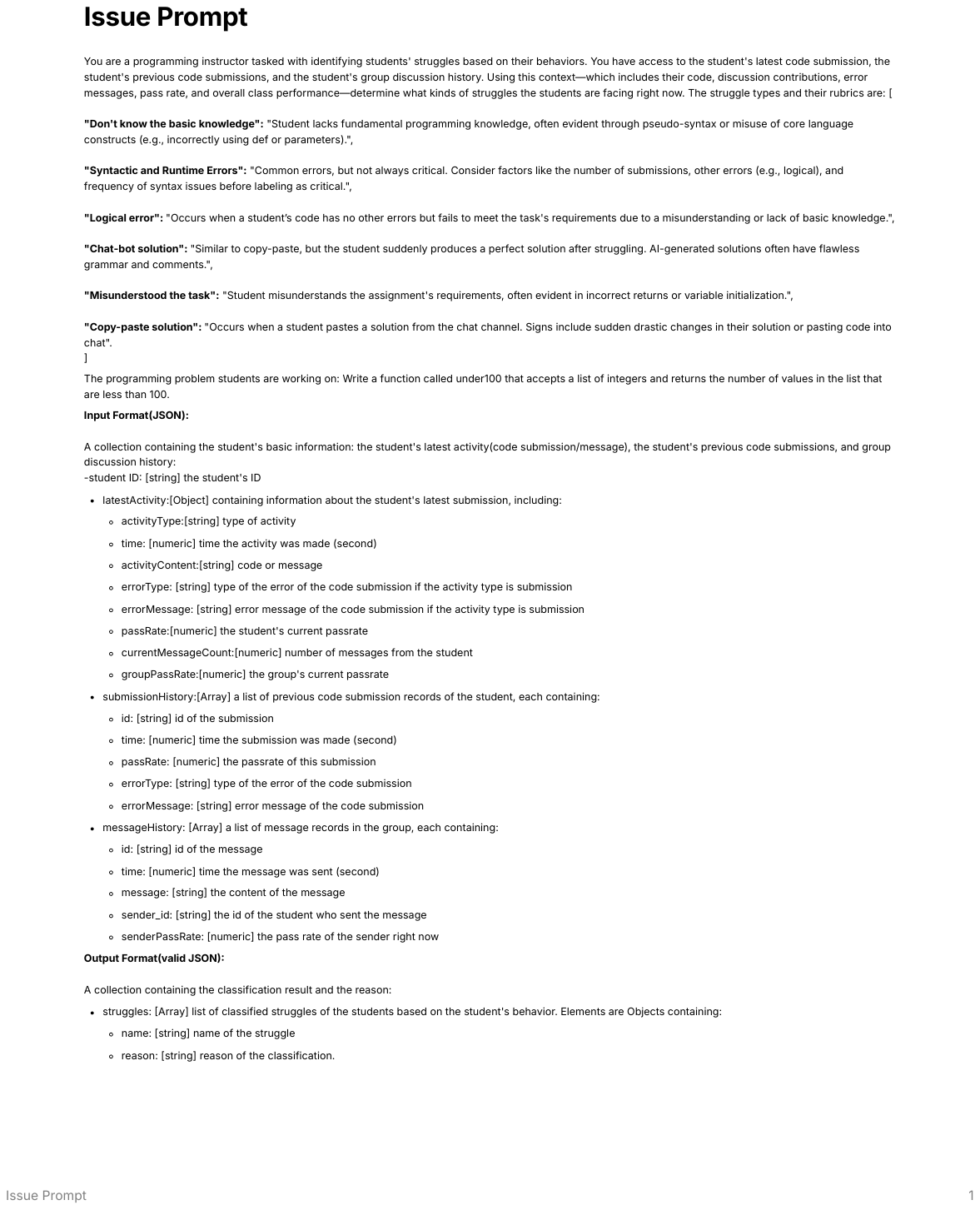}
    \caption{Prompt for identifying student's struggles 1}
    \label{fig:struggle_1}
\end{figure*}
\newpage
\begin{figure*} [!h]
    \centering
    \includegraphics[width=0.91\linewidth]{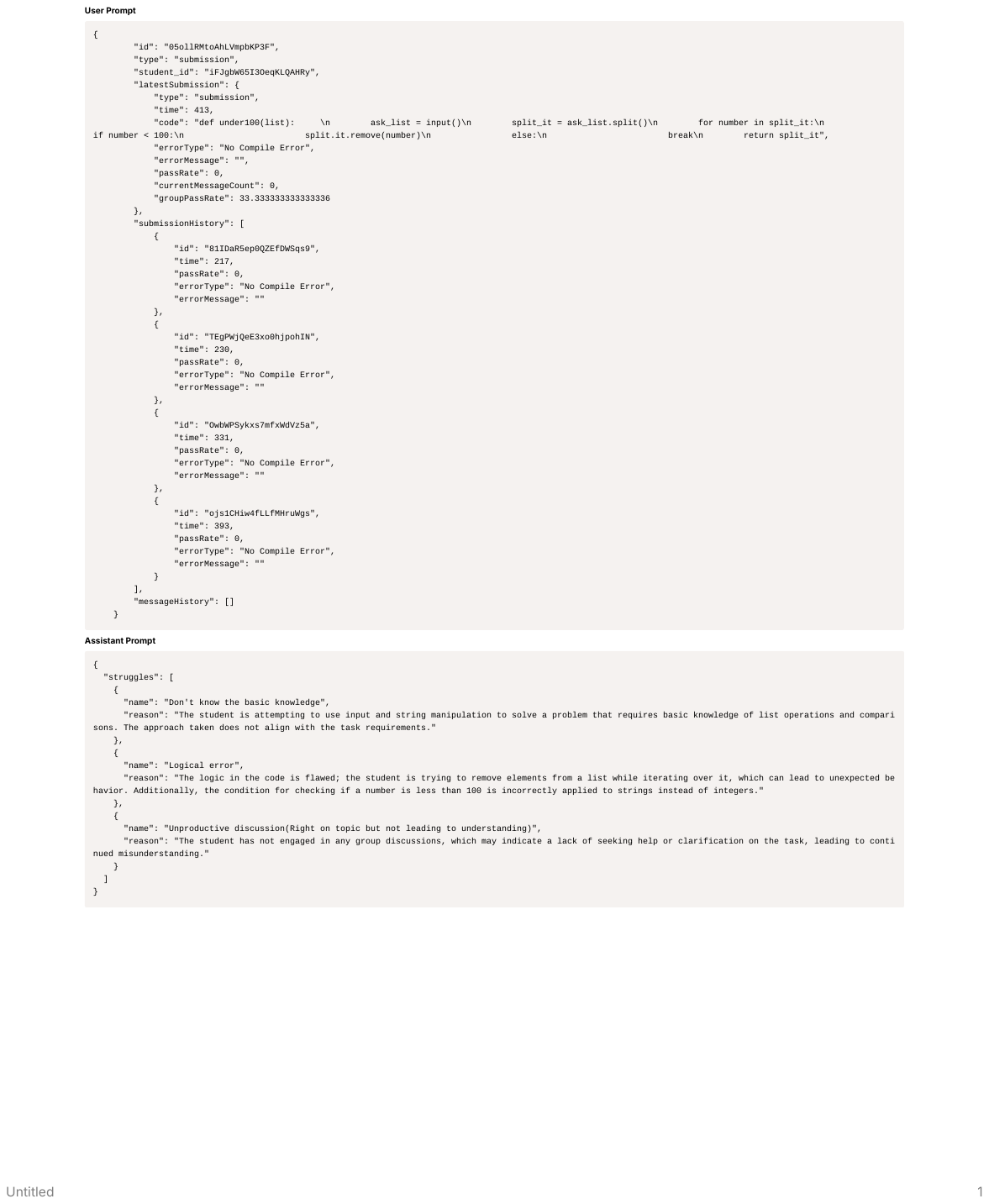}
    \caption{Input and Output Example for identifying student's struggle 1}
    \label{fig:sturggle_2}
\end{figure*}
\newpage
\begin{figure*} [!h]
    \centering
    \includegraphics[width=0.94\linewidth]{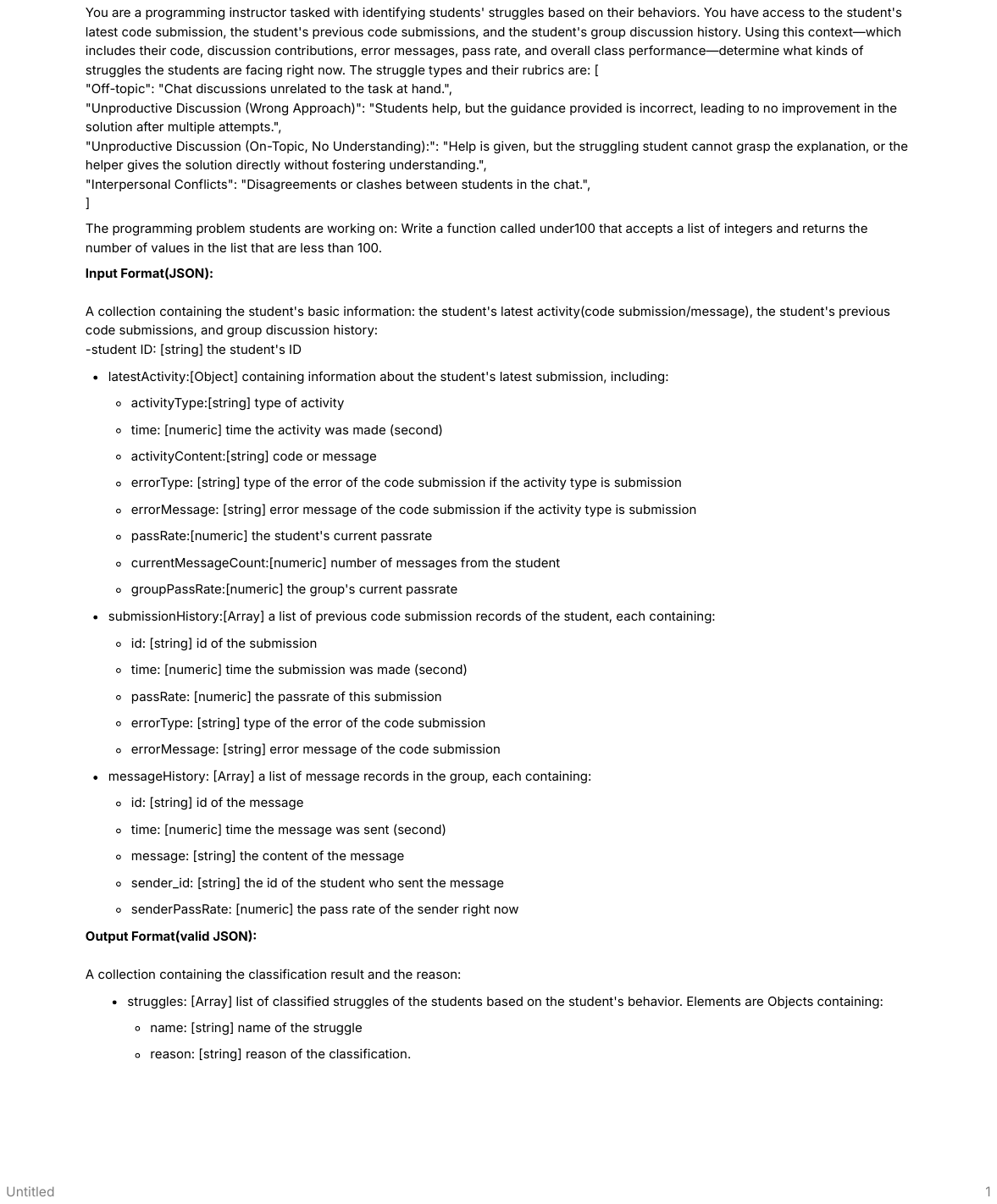}
    \caption{Prompt for identifying student's struggles 2}
    \label{fig:struggle_3}
\end{figure*}
\newpage
\begin{figure*} [!h]
    \centering
    \includegraphics[width=0.94\linewidth]{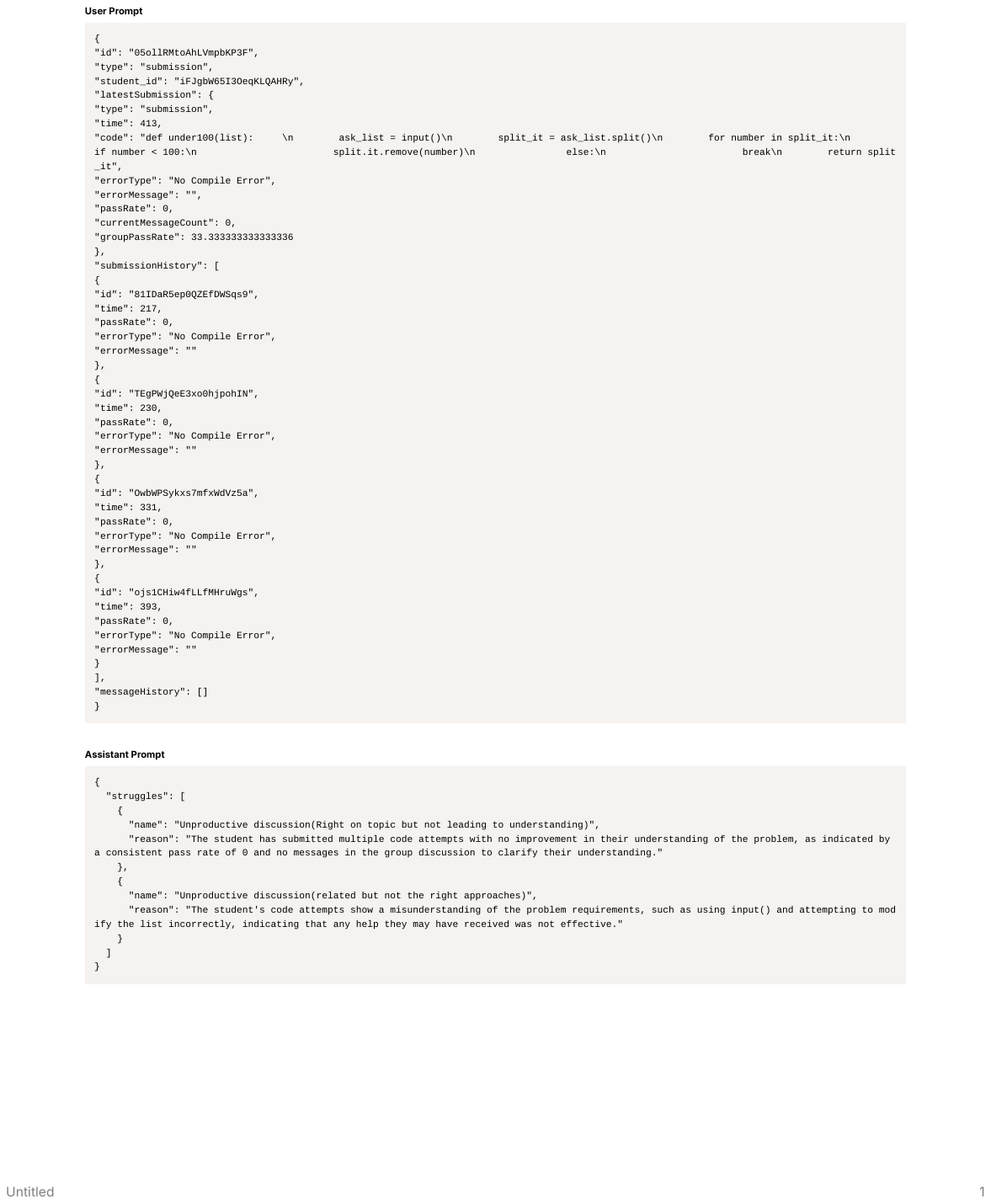}
    \caption{Input and Output Example for identifying student's struggle 2}
    \label{fig:struggle_4}
\end{figure*}
\newpage
\subsection{Appendix E: Prompt for generating personalized student feedback}~\label{appendixE}

\begin{figure*} [!h]
    \centering
    \includegraphics[width=0.9\linewidth]{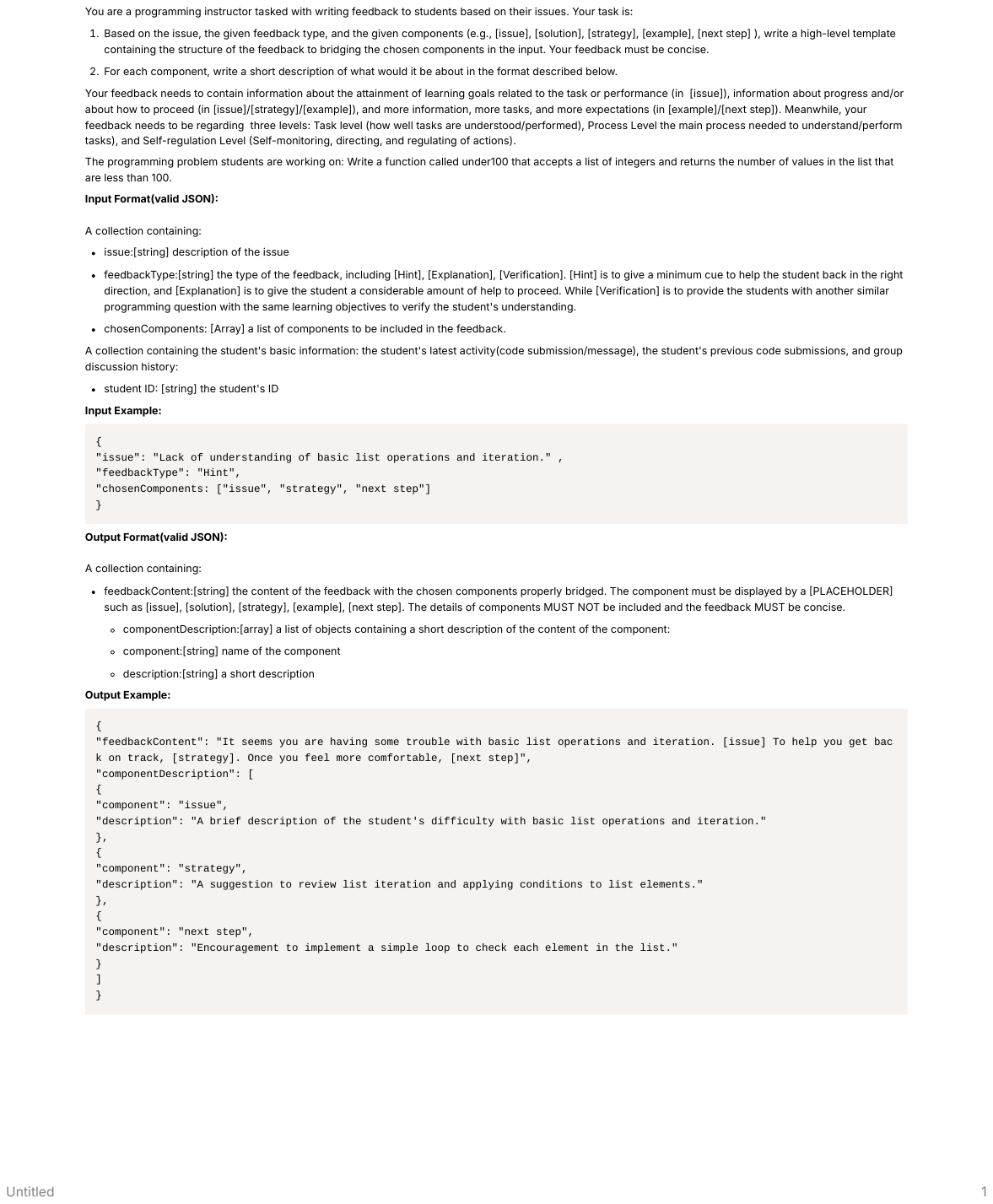}
    \caption{Prompt for generating feedback templates}
    \label{fig:feedback_1}
\end{figure*}
\newpage
\begin{figure*} [!h]
    \centering
    \includegraphics[width=0.9\linewidth]{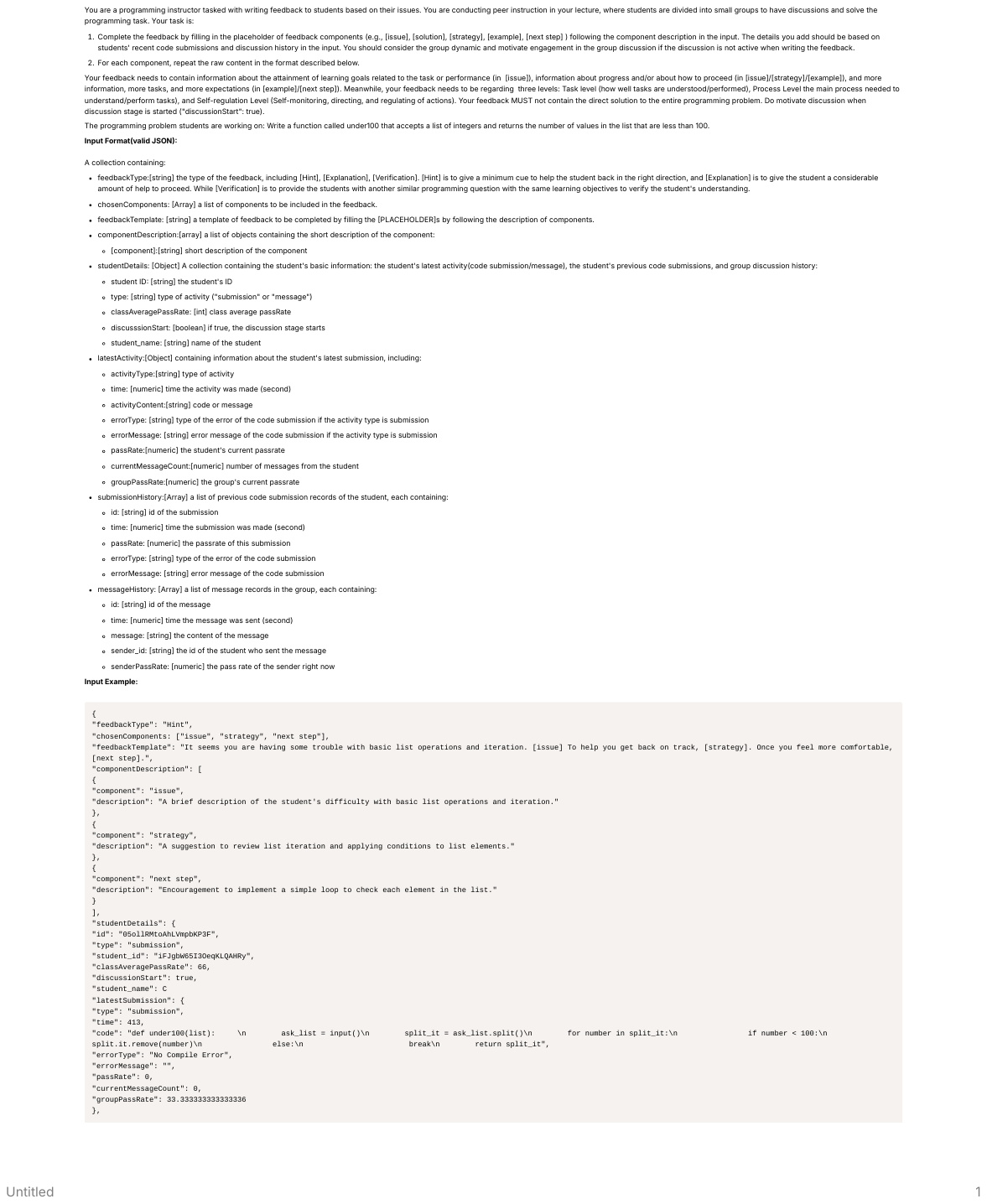}
    \label{fig:feedback_2}
\end{figure*}
\newpage
\begin{figure*} [!h]
    \centering
    \includegraphics[width=0.9\linewidth]{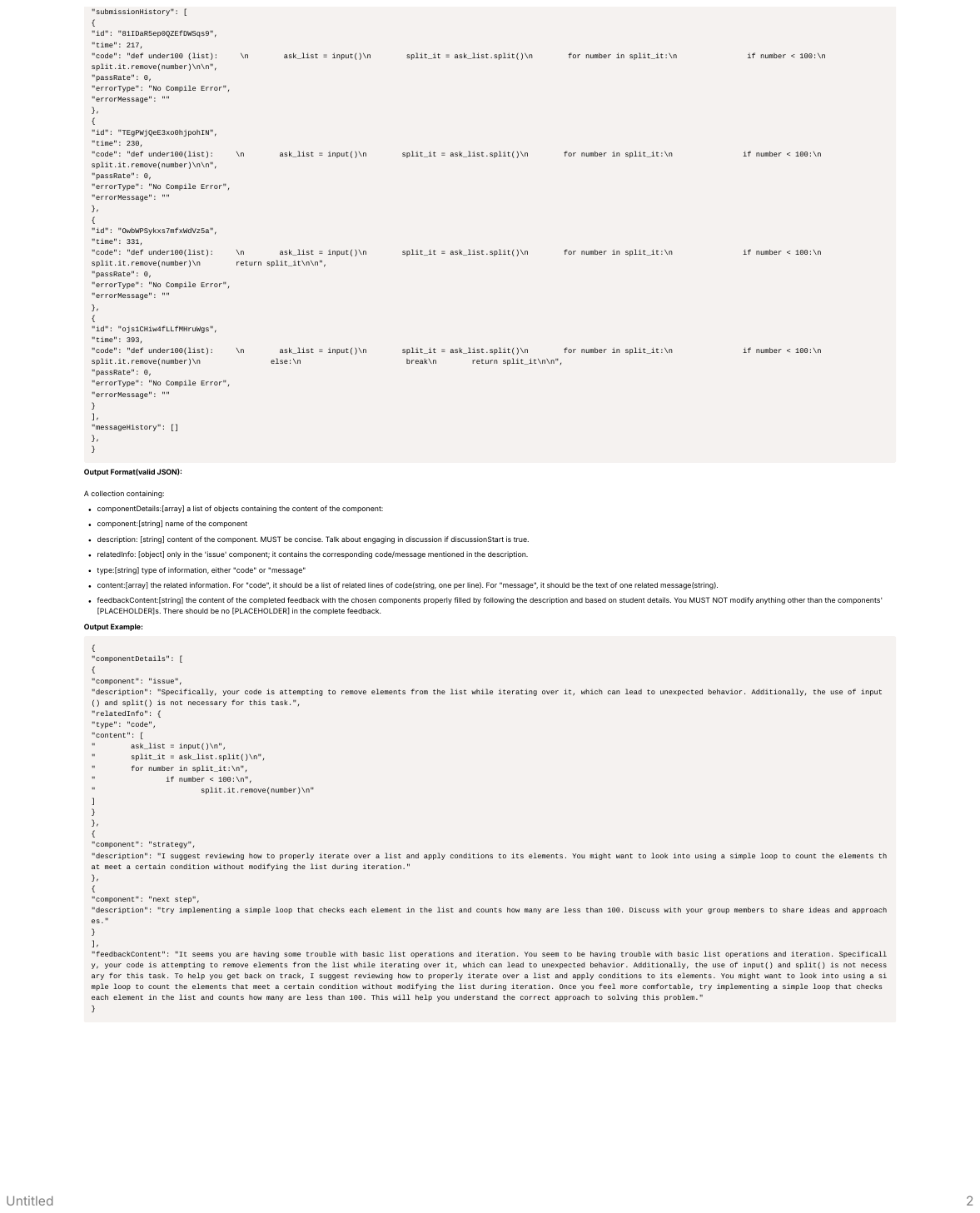}
    \caption{Prompt for generating feedback content}
    \label{fig:feedback_3}
\end{figure*}
\newpage

\subsection{Appendix F: Labeling Rubric}~\label{appendixF}
\begin{table*}[!h]

\caption{Rubric for annotating student coding and conversation data}
\centering
\renewcommand{\arraystretch}{1.5} % Increase row height
\setlength{\tabcolsep}{10pt} % Adjust column separation
\label{table:issuetaxonomy}
\begin{adjustbox}{max width=\textwidth}
\begin{tabular}{ >{\centering\arraybackslash}m{3cm}  p{12cm} }
\hline
\textbf{Issue} & \textbf{Definition} \\
\hline
Copy-Paste Solution & This critical situation occurs when a student pastes a solution from another source. \\
\hline
Don’t know the basic knowledge & In this situation, the current student lacks the necessary knowledge of the programming language and its concepts. Most commonly in the form of typing pseudo-syntax (syntax that they’ve made up to accomplish the task, or combining existing syntax in invalid ways). \\
\hline
Syntax error & Syntax errors that be evaluated based on factors such as submission frequency, coexistence with other issues, and error quantity. \\
\hline
Logical error & Flaws in the student’s code that prevent them from passing all the test cases. Usually, these flaws stem from a misunderstanding of the problem or lack of basic knowledge. \\
\hline
Chat-bot solution & Look for moments where there are no solutions in the chat from any other students, and the current student will produce a perfect solution after visibly struggling for some time. \\
\hline
Off-topic & This critical situation occurs when there are ongoing discussions in the chat that are not related at all to the current task. \\
\hline
Unproductive discussion (related but not the right approaches) & This type of scenario occurs when students try to help each other, but it does not result in success or change in the student’s solution. \\
\hline
Unproductive discussion (Right on topic but not leading to understanding) & Multiple attempts of students helping each other, but the student cannot grasp an understanding regardless of if the other student is explaining it perfectly or not. Or the student trying to help gives out the solution directly, without aiding the struggling student in understanding the concept. \\
\hline
Interpersonal Conflicts & When there is disagreement among the students in chat. Clashes between students. \\
\hline
Misunderstood the task & This critical situation happens when the student fails to grasp the requirements or objectives of the assignment. \\
\hline
\end{tabular}
\end{adjustbox}
\end{table*}
\pagebreak
\subsection{Appendix G: Rubric for annotating feedback}~\label{appendixG}
\textbf{Incorrect Feedback (-1):}

\textit{Characteristics:}
\begin{enumerate}
    \item The feedback is factually incorrect, wrong, or misleading.
    \item Provides direct answers without explanation, preventing learning or understanding.
    \item Uses confusing or unclear language, making it hard for students to apply the feedback.
    \item Offers overly negative or harsh criticism without actionable steps for improvement.
    \item Contains errors that can misguide students or reinforce incorrect understanding of concepts.
\end{enumerate}
\textit{Impact:}
\begin{enumerate}
    \item This type of feedback leads to confusion, frustration, or lack of confidence in the student. It may also result in students applying incorrect practices.
\end{enumerate}

\textbf{Shallow Feedback (0):}
\textit{Characteristics:}
\begin{enumerate}
    \item The feedback is factually correct but lacks depth or clarity.
    \item Offers minimal insight or vague suggestions, not addressing specific issues with enough detail.
    \item Does not guide the student toward improvement or deeper understanding.
    \item Is clear but not particularly encouraging, leaving the student without a clear sense of next steps.
Avoids technical errors, but lacks motivation or educational value.
\end{enumerate}
\textit{Impact:}
\begin{enumerate}
    \item This type of feedback may leave the student feeling unsure of how to proceed or improve. While technically correct, it doesn't significantly help the student grow.
\end{enumerate}

\textbf{High-quality Feedback (1):}
\textit{Characteristics:}
\begin{enumerate}
    \item The feedback is clear, specific, and actionable, helping the student understand both their strengths and areas for improvement.
    \item Encourages learning by explaining why something is incorrect and offering suggestions for how to fix it.
    \item Provides context for why a certain approach is better and explains the concepts behind it.
    \item Balances criticism with encouragement, helping build the student’s confidence while addressing errors.
    \item Links feedback to learning objectives, reinforcing key programming concepts in a positive and constructive manner.
\end{enumerate}
\textit{Impact:}
\begin{enumerate}
    \item High-quality feedback helps students understand what they did right, where they went wrong, and how to improve. It fosters a growth mindset and promotes deeper learning, building both technical and problem-solving skills.

\end{enumerate}

% \appendix

\end{document}